%% file: smec-paper.tex
\documentclass[letterpaper,twocolumn,10pt]{article}
\usepackage{usenix-2020-09}
\usepackage{kim-paper-macros}
\renewcommand{\arraystretch}{1.5}

\hypersetup{pdfstartview=FitH,pdfpagelayout=SinglePage}

\newcommand{\modified}[1]{{\color{black}#1}}

\usepackage[available,functional,reproduced]{usenixbadges}

\begin{document}

\newcommand{\sys}{\mbox{{SMEC}}\xspace}
\newcommand{\dallas}{\mbox{{Dallas}}\xspace}
\newcommand{\nanjing}{\mbox{{Nanjing}}\xspace}
\newcommand{\seoul}{\mbox{{Seoul}}\xspace}
\newcommand{\tutti}{\mbox{{Tutti}}\xspace}
\newcommand{\arma}{\mbox{{ARMA}}\xspace}

\pagenumbering{gobble}

\title{Enabling SLO-Aware 5G Multi-Access Edge Computing with \sys}

\author{
    {\rm Xiao Zhang  \quad Daehyeok Kim}\\
    The University of Texas at Austin\\
}

\maketitle
\frenchspacing
\input{00_abstract}
\pagestyle{empty}

\input{01_intro}
\input{02_motivation}
\input{03_overview}
\input{04_ran_sched}
\input{05_edge_sched}
\input{06_impl}
\input{07_eval}
\input{08_discuss}
\input{09_concl}
\input{acks}
\label{EndOfPaper}

{
\bibliographystyle{plain}
\bibliography{ref}
}

\input{10_appendix.tex}

\end{document}

%% file: 00_abstract.tex
\begin{abstract}
Multi-access edge computing (MEC) promises to enable latency-critical applications by bringing computational power closer to mobile devices, but our measurements on commercial MEC deployments reveal frequent SLO violations due to high tail latencies.
We identify resource contention at the RAN and the edge server as the root cause, compounded by SLO-unaware schedulers.
Existing SLO-aware approaches require RAN--edge coordination, making them impractical for deployment and prone to poor performance due to coordination delays, limited heterogeneous application support, and ignorance of edge resource contention.
This paper introduces \sys{}, a practical, SLO-aware resource management framework that facilitates deadline-aware scheduling through fully decoupled operations at the RAN and edge servers.
Our key insight is that standard 5G protocols and application behaviors naturally provide information exploitable for SLO-aware management without extensive infrastructure or application changes.
Evaluation on our 5G MEC testbed shows that \sys{} achieves 90--96\% SLO satisfaction versus under 6\% for existing approaches, while reducing tail latency by up to 122$\times$.
We have open-sourced \sys{} at \url{https://github.com/smec-project}.
\end{abstract}

%% file: 01_intro.tex
\section{Introduction}
\label{sec:intro}

Multi-access edge computing (MEC) brings computational power closer to mobile devices by connecting with 5G cellular networks~\cite{wavelength, tmobile}.
It enables latency-critical (LC) applications, from smart stadium~\cite{www-orange-5gstadium,www-red5perfect5G-in-venue} and AR/VR~\cite{www-cloudxr,liu2019edge} to cloud gaming~\cite{www-xbox-cloud,www-ericsson-cloud-gaming} and autonomous driving~\cite{www-dt-driving} to offload their compute-intensive tasks to edge servers. 
These applications typically operate through request-response interactions between clients and edge servers, where each request and response may span multiple packets, and each application must meet strict service-level objectives (SLOs) on request-to-response latency.

Unfortunately, today's MEC deployments fall short of this promise.
Our measurement studies across three cities from three countries reveal that commercial MEC services suffer from high tail latencies that frequently violate application SLOs (\autoref{sec:motivation}).
The root cause is resource contention at both the radio access network (RAN) and edge servers, compounded by resource schedulers that lack SLO-awareness, causing resource allocation decisions to directly impact request progress and SLO satisfaction.

Existing SLO-aware scheduling approaches suffer from practical limitations that prevent real-world deployment.
Systems like \tutti{}~\cite{xu2022tutti} and \arma{}~\cite{arma:mobisys25} require explicit coordination between RAN schedulers and edge applications, which is impractical because RAN infrastructure and edge servers are typically managed by different entities (\eg Verizon operates the RAN while AWS provides edge compute).
Even if such coordination were feasible, we find that feedback delays between edge servers and RAN prevent timely resource allocation when applications need it most. 
Moreover, these solutions focus narrowly on specific application types while ignoring edge compute resource management, failing to address the heterogeneous nature of MEC workloads where multiple applications with diverse SLO requirements compete for both network and compute resources.

This paper introduces \sys{},\footnote{Short for \textbf{S}LO-aware \textbf{MEC}} a practical SLO-aware resource management framework that operates through complete decoupling. 
\sys{} runs entirely independent resource managers at the RAN and edge server, each making deadline-aware resource allocation decisions without any coordination between them. 
Both managers prioritize latency-critical applications approaching their deadlines by estimating \emph{remaining time budgets}, enabling timely resource allocation when applications need it most.
This decoupled design enables deployment in realistic settings where different entities control the RAN and edge infrastructure.
\sys{} supports heterogeneous workloads, ensuring LC tasks meet their deadlines without starving best-effort (BE) traffic.

While our approach sounds promising, designing such a decoupled framework introduces three key technical challenges.
First, the RAN scheduler must identify application request boundaries to estimate time budgets without packet payload inspection due to strict timing constraints.
Second, the edge server must estimate the uplink transmission time already consumed by incoming requests and predict the future downlink transmission time for responses, all without visibility into RAN delays.
Third, the edge server also must predict processing time across heterogeneous workloads without requiring intrusive application modifications.

Our key insight is that standard 5G protocols and applications' request-response patterns provide useful signals to solve these challenges independently.
We find that these signals can be exposed with minimal or no application modifications.
First, we exploit patterns in 5G control signals (\eg Buffer Status Reports) to infer new request arrivals at the RAN without payload inspection.
Second, at the edge, we leverage the stability of downlink transmissions to estimate network latency via a lightweight probing protocol and a client-side API without coordinating with the RAN.
Third, for processing time estimation at the edge, we exploit key lifecycle events of requests exposed through a server-side API, enabling \sys{} to track execution history and predict processing times for incoming requests.
Based on this information, \sys{} enables the RAN and the edge to compute the remaining time budget for requests independently and manage resources based on these budgets.    

\modified{We implemented \sys{} as user-space resource managers in C++ and Python that run across the RAN, edge servers, and client devices.
At the RAN, our resource manager operates as a pluggable scheduling module for srsRAN's MAC layer, implementing request identification and deadline-aware scheduling without affecting other RAN functionalities.
At the edge, our resource manager runs as a user-space daemon that estimates network and processing times while dynamically managing CPU and GPU allocations.
Client devices run a lightweight timing daemon to support the probing protocol for network latency estimation.
We have open-sourced \sys{}, including the evaluated applications and experiment scripts: \url{https://github.com/smec-project}.}

We evaluated \sys{} on our private 5G MEC testbed using three LC applications and one BE application running on 12 client devices.
\sys{} achieves 90--96\% SLO satisfaction under both static and dynamic workloads, compared to less than 6\% for existing approaches.
It reduces tail latency by up to 122$\times$ for uplink-intensive applications and consistently improves P99 latency by 2--89$\times$ across all workloads.
\modified{Importantly, \sys{} allows  BE applications to fairly share remaining bandwidth without prolonged starvation.}

%% file: 02_motivation.tex
\section{Background and Motivation}
\label{sec:motivation}

\subsection{Primer on Multi-Access Edge Computing}
\label{sec:primer}

Multi-Access Edge Computing (MEC) brings computational resources closer to end users by deploying compute infrastructure close to cellular base stations~\cite{etsi-mec}.
The MEC architecture consists of three key components: the RAN that manages wireless spectrum and radio resource scheduling, edge servers that provide CPU and GPU resources, and the core network that connects the RAN and the servers.
In the typical MEC request-response model, LC applications offload compute-intensive tasks from client devices to edge servers.
Clients send requests via RAN uplink, edge servers process them locally, and responses return via RAN downlink.

\begin{table}[]
\setlength{\tabcolsep}{4pt}
\renewcommand{\arraystretch}{1.0}
\footnotesize
\centering
\begin{tabular}{@{}p{2.1cm}p{2.0cm}p{0.8cm}p{1.2cm}p{1.2cm}@{}}
\toprule
\textbf{Applications} & \textbf{Offloaded Task} & \textbf{SLO} & \textbf{UL/DL Load} & \textbf{Compute Resource} \\ \midrule
Smart stadium~\cite{www-orange-5gstadium} & Video transcoding & 100ms & High/High & CPU \\
Augmented reality~\cite{mao2019delay, padmanabhan2023gemel} & Object detection & 100ms & Med/Low & GPU \\
Video conferencing~\cite{www-zoom-latency, www-wikipedia-audio-latency} & Super resolution & 150ms & Low/High & GPU \\ \bottomrule
\end{tabular}
\caption{Examples of MEC applications evaluated in this paper, each with distinct SLO, network (uplink/downlink) load, and compute resource requirements. Details are described in~\autoref{sec:eval-setup}.}
\label{tbl:apps}
\vspace{-1em}
\end{table}

\autoref{tbl:apps} shows the MEC applications we focus on in this paper, each with distinct SLO, network load, and compute resource requirements.
We target realistic MEC deployments where multiple user equipment (UE) devices (\eg smartphones, cameras, AR headsets) simultaneously offload compute-intensive tasks to edge servers.

However, as we demonstrate next, the reality of current MEC deployments falls short of these performance promises.

\subsection{Unpredictable Performance of MEC}
\label{sec:unpredictable}

We benchmarked MEC deployments across three cities in the US, South Korea, and China (Dallas, Seoul, and Nanjing) to quantify this performance gap.

\mypara{Measurement setup.}
We deployed the smart stadium and augmented reality applications from~\autoref{tbl:apps} using a laptop client connected to a 5G smartphone hotspot and edge VMs provisioned through edge service providers (\eg AWS in the US).
Each application runs in isolation (\ie no contention on the VM), generates 10,000 requests, and we measure end-to-end latency from request transmission to complete response reception.
For brevity, we focus our analysis on the smart stadium application, presenting results from all three cities that employ different combinations of cellular operators and cloud providers.
We observe similar trends for the augmented reality application across all cities (\autoref{appnd:ar}).

\begin{figure}[]
    \centering
    \includegraphics[width=\columnwidth]{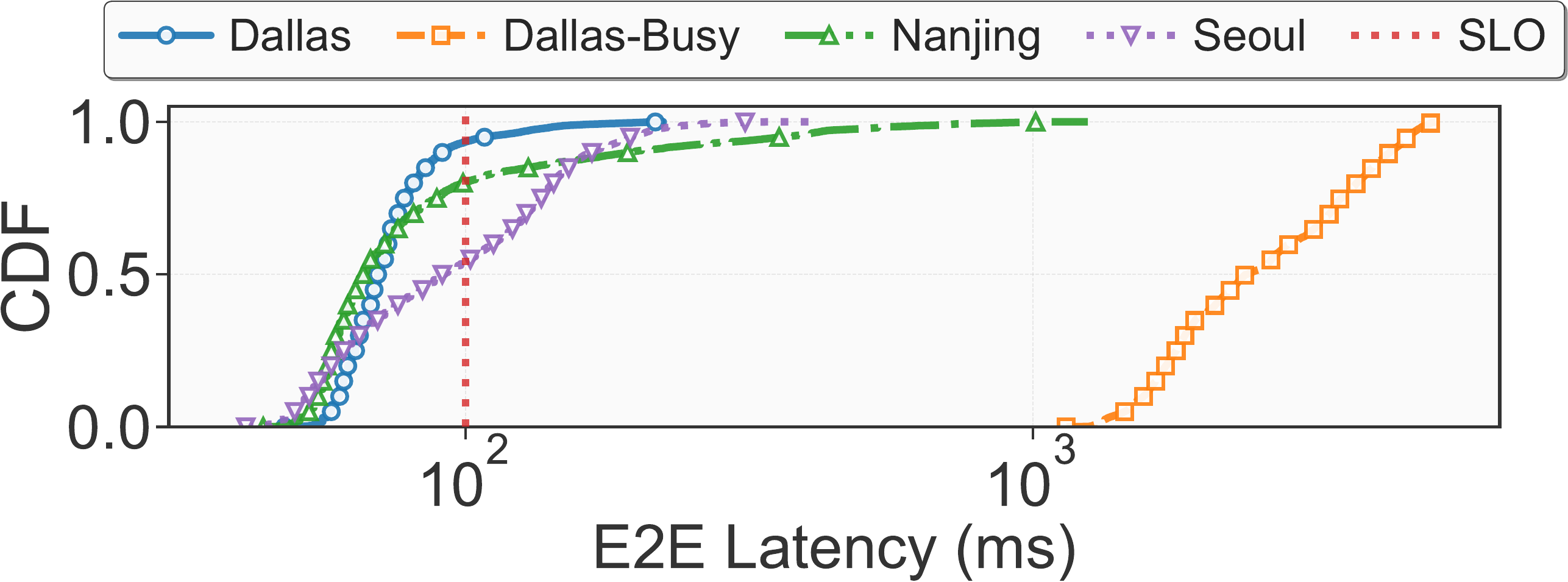}
    \tightcaption{End-to-end latency for the smart stadium application without edge resource contention across MEC deployments in three cities. The dotted red line indicates the SLO.}
    \label{fig:motive-e2e-transcoding-ran}
\end{figure}

\mypara{Results.}
\autoref{fig:motive-e2e-transcoding-ran} reveals unpredictable performance characterized by high tail latencies across all deployments, even without compute resource contention on the edge server.
Specifically, 7\%, 20\% and 47\% of requests exceed their SLO requirements in \dallas, \nanjing, and \seoul, respectively, during low network activity periods (measured at 2am).
Although median latencies remain below the SLO threshold, the P95 and P99 latencies are substantially higher, resulting in an inconsistent user experience.
This problem intensifies under higher network load: when additional UE devices access the 5G network and increase contention for RAN resources (\dallas-Busy), even the median latency exceeds the SLO requirements.

\subsection{Root Causes of Unpredictability}
\label{sec:root-cause}

To understand the sources of this unpredictability, we investigate what contributes to the tail latency in both the RAN and the edge server.

\subsubsection{RAN Resource Contention}
\label{sec:wireless-bottlenecks}

To analyze RAN-induced latency variability, we decompose end-to-end latency into uplink and downlink transmission phases.
Using Precision Time Protocol (PTP)~\cite{ptp-std} for clock synchronization between client device and edge server via a stable out-of-band wired connection, we developed a synthetic application that  measures uplink and downlink latency separately while varying request and response sizes.
Specifically, we measured the time to receive complete requests at the server (uplink) and complete responses at the client (downlink) while varying data size. 

\begin{figure}[]
    \centering
    \includegraphics[width=\columnwidth]{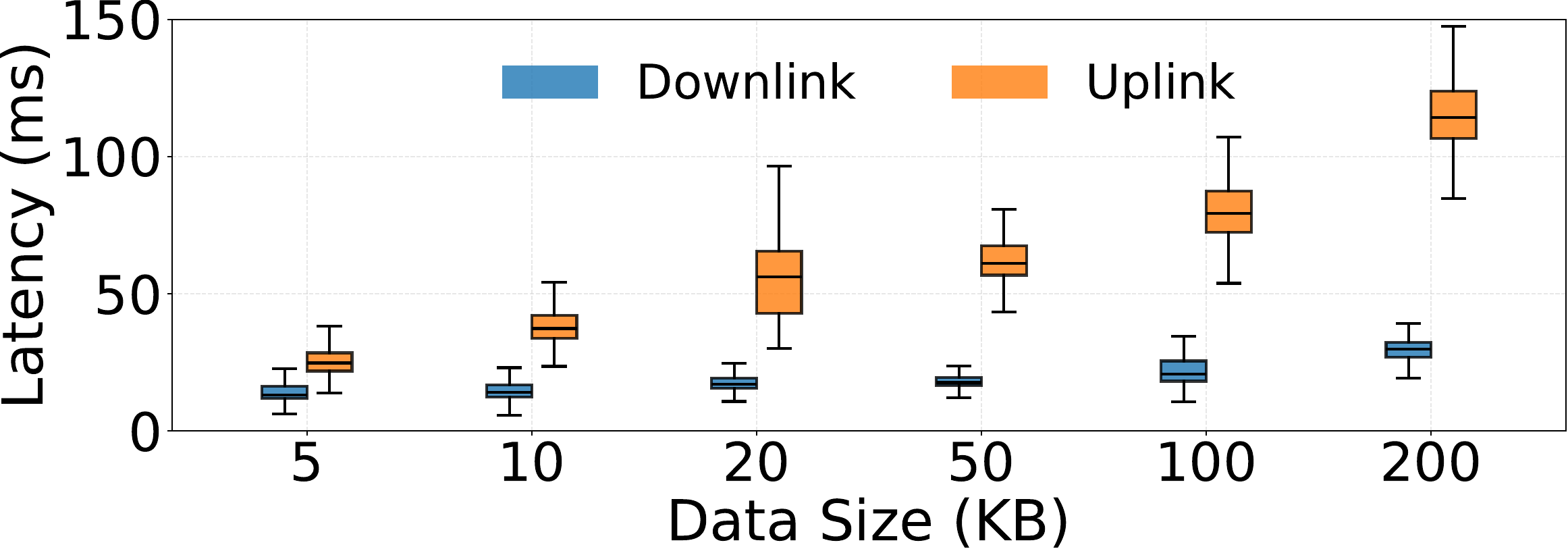}
    \tightcaption{Network latency variability for uplink and downlink transmissions across different data sizes in \dallas.
 Data size refers to request size for uplink and response size for downlink.}
    \label{fig:motive-decompose}
\end{figure}

\autoref{fig:motive-decompose} reveals an asymmetry that explains MEC's unpredictable performance: 
uplink latency exhibits high variability, especially for larger request sizes, while downlink latency remains stable.
This reflects how cellular networks provision fewer uplink slots than downlink slots, causing higher contention for uplink transmissions.
We observe similar trends across all measured cities (\autoref{appnd:latency-variability}).

To understand the specific mechanisms causing uplink jitter, we leveraged our srsRAN 5G~\cite{srsran}-based testbed for in-depth analysis (details in~\autoref{sec:eval-setup}), since we have no visibility into the internals of public cellular networks.
The srsRAN stack employs the proportional fair (PF) scheduling algorithm~\cite{kelly1997charging,jalali2000data} used in commercial deployments, allowing us to emulate real-world scheduling behavior.

We ran the smart stadium application on the testbed and monitored MAC layer state by collecting buffer status reports (BSRs) sent from each UE.
As a background workload, we also deployed five file transfer UEs.
BSRs indicate remaining data in UE-side transmission buffers, providing direct insight into whether the scheduler allocates sufficient RAN resources\footnote{We use the term RAN resources to refer to uplink and downlink spectrum resources managed by the MAC scheduler (\ie Physical Resource Blocks in 5G).} to meet application demands.

\begin{figure}[]
    \centering
    \includegraphics[width=\columnwidth]{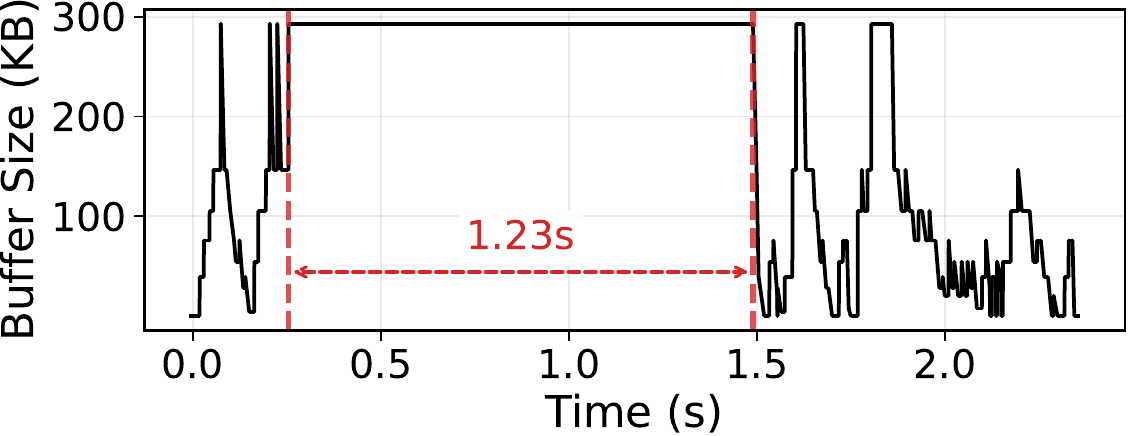}
    \tightcaption{
        Smart stadium UE's uplink buffer status changes over time. 
        300 KB is the maximum for BSR from UE to the RAN, which means UE may buffer more than 300 KB.
    }
    \label{fig:uplink-bsr}
\end{figure}

\autoref{fig:uplink-bsr} shows persistent non-zero BSR (>1s), indicating uplink starvation from PF scheduling.
This resource starvation directly translates to the observed uplink jitter; inadequate resource allocation causes unpredictable buffering delays that lead to SLO violations.

The root cause is PF schedulers' lack of SLO awareness.
They prioritize UEs with relatively better channel conditions to their historical average throughput, balancing fairness and efficiency without considering SLO requirements.
When multiple UEs compete for uplink resources, LC applications cannot obtain timely allocations, missing SLO requirements even when aggregate bandwidth is sufficient.

\begin{figure}[]
    \centering
    \includegraphics[width=\columnwidth]{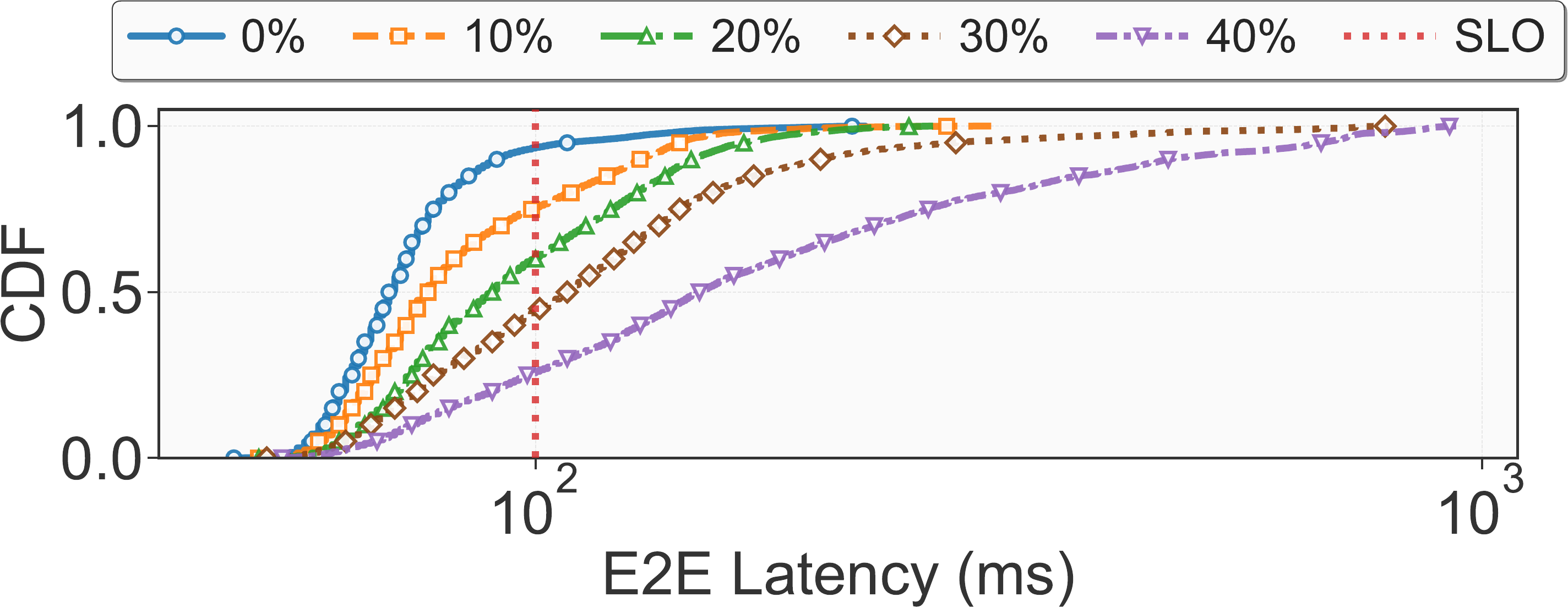}
    \tightcaption{End-to-end latency for smart stadium under different levels of compute resource contention in \dallas. The dotted red line indicates the SLO.}
    \label{fig:motive-e2e-transcoding-compute}
\end{figure}

\subsubsection{Edge Compute Resource Contention}
\label{sec:compute-bottlenecks}

While wireless resource allocation causes unpredictable uplink performance, edge compute resources present an additional bottleneck in MEC deployments.
To quantify the impact of compute resource contention, we emulate competing offloaded tasks by running a CPU stressor (stress-ng~\cite{stress-ng}) with varying CPU usage levels.
\autoref{fig:motive-e2e-transcoding-compute} shows that compute resource contention significantly contributes to unpredictable performance of the smart stadium application.
As CPU load increases, tail latencies grow substantially, confirming that inadequate management of edge compute resources causes processing delays that cascade into SLO violations.
This problem extends to GPU-intensive applications and other MEC deployments (\autoref{appnd:compute-contention}).

\subsection{Prior Approaches and Limitations}
\label{sec:prior}

Having identified the root causes of unpredictable MEC performance, we now examine existing resource management approaches and discuss why they fall short for our setting.

\mypara{MEC-specific resource management.}
Recent MEC resource management proposals suffer from impractical deployment requirements and limited scope as we show in~\autoref{sec:eval}.
\tutti{}~\cite{xu2022tutti} assumes explicit coordination between the RAN  and edge applications, requiring servers to notify the RAN upon receiving requests and transport protocol modifications to identify ``first packets.''
This is impractical because RAN infrastructure and edge servers are typically managed by different entities (\eg Verizon operates the RAN while AWS provides edge compute).
Even setting deployment issues aside, \tutti{} fails to timely accelerate LC requests due to feedback delays between edge servers and RAN.
Moreover, \tutti{} only supports homogeneous applications with identical SLOs.
\arma{}~\cite{arma:mobisys25} similarly requires coordination between edge servers and RAN schedulers, is narrowly tailored to video analytics, and allows non-LC applications to block LC ones when their uplink bandwidth usage is high.
Both systems focus solely on RAN scheduling while ignoring edge compute resource management for co-existing heterogeneous applications.

\mypara{SLO-aware resource management in clouds.}
A natural approach for managing compute resources is to adapt existing SLO-aware resource management solutions from cloud environments.
However, these solutions are ill-suited for our target LC applications.
PARTIES~\cite{parties:asplos19} reactively adjusts server resource partitions based on SLO feedback from clients, but fails in MEC where wireless feedback delays mean multiple requests miss deadlines before adjustments take effect.
Caladan~\cite{caladan:osdi20} employs proactive CPU core allocation without requiring client feedback but requires extensive application modifications.
ML-based schedulers like OSML~\cite{liu2023intelligent} incur inference overhead and operate at second-level intervals, making them too slow to react within the tens of milliseconds required for LC applications.
Similarly, UFO~\cite{peng2024ufo} relies on changes in system-wide metrics (\eg CPU utilization, scheduling frequency) captured at second-level intervals to infer potential SLO violations and makes scheduling decisions, which reacts too slowly for LC applications. 
Vessel~\cite{lin2024fast} assumes all processes share one memory space and relies on hardware-assisted core allocation, limiting its applicability in multi-tenant MEC settings and beyond CPU scheduling.

\mypara{End-to-end rate control mechanisms.}
End-to-end congestion control mechanisms~\cite{carlucci2017congestion,zhu2020network,johansson2017self} are ill-suited for LC applications in MEC environments.
Congestion control requires hundreds of milliseconds to converge, causing SLO violations before it can react to uplink congestion.
Also, the 5G uplink channel quality fluctuates rapidly due to limited UE transmission power and varying user counts, making it nearly impossible for congestion control to stabilize.
\modified{Even approaches that leverage 5G control-channel information (\eg PBE-CC~\cite{xie2020pbe}) to react quickly run solely at the end host and cannot directly influence RAN uplink scheduling.
Under severe wireless contention, the RAN may allocate minimal or no uplink resources to a UE, so sender-side adaptation alone is insufficient to satisfy SLO requirements.}
Additionally, many LC applications rely on constant or variable bitrate encoding to maintain video quality~\cite{Fortinet2020_IPSurveillanceCameraBandwidth,Sony_BRCX400_Specs}, where reactive rate adaptation directly degrades quality.

\modified{\mypara{URLLC.}
5G Ultra-Reliable Low-Latency Communication (URLLC), a service category designed for mission-critical applications, achieves low latency and high reliability through conservative radio configurations and dedicated resource reservations~\cite{esti-5g-arch}.
However, this approach reduces spectrum efficiency: reserved resources often remain underutilized, and reliability-enhancing techniques consume additional radio resources without commensurate benefit.
The inefficiency is particularly pronounced for uplink-heavy workloads where resource demand is dynamic, making static reservations ill-suited for our setting.}

%% file: 03_overview.tex
\section{Overview of \sys}
\label{sec:overview}

We now present \sys{}, a practical SLO-aware resource management framework for MEC that support multiple heterogeneous applications through a decoupled scheduling approach.

\subsection{Design Goals}
\label{sec:design-goals}

To address the limitations identified in~\autoref{sec:prior}, we aim to design \sys{} around four goals:

\mypara{G1: No coordination between RAN and edge servers.}
\sys{} should operate with completely decoupled schedulers at the RAN and edge servers. 
This will address the practical deployment and technical challenges that make coordination-based approaches infeasible.

\mypara{G2: Compatibility with existing infrastructure and applications.}
\sys{} should require no significant modifications to existing 5G stacks, edge servers, or applications.
This will enable incremental deployment without disrupting existing MEC infrastructures or requiring application re-engineering.

\mypara{G3: Resource management for heterogeneous applications.}
\sys{} should provide resource allocation across RAN and edge resources for multiple competing applications with diverse SLOs.
This will enable MEC deployments where heterogeneous applications must coexist efficiently.

\mypara{G4: SLO satisfaction through deadline-aware resource scheduling.}
\sys{} should prioritize SLO satisfaction over conventional fairness objectives that often lead to latency variability. 
Scheduling decisions must account for application deadlines and current resource availability, ensuring LC applications obtain the resources they need while still preventing starvation of BE  applications.

\subsection{Challenges}
\label{sec:challenges}

Realizing SLO-aware scheduling without RAN--edge coordination presents three challenges:

\mypara{C1: New request identification at the RAN.}
RAN protocol layers, especially the MAC layer where resource allocation decisions are made, lack visibility into application payloads and find it impractical to parse application data due to the tight timing requirements of RAN processing.
This makes it challenging to identify when new requests begin and trigger appropriate SLO-aware scheduling decisions.

\mypara{C2: Network latency estimation at the edge.}
The edge server needs to estimate the network transmission latency that each request has consumed during uplink transmission and will consume during downlink transmission for responses to compute the remaining time budget before the SLO deadline expires.
However, it has no visibility into these transmission delays, making SLO-aware resource allocation challenging.

\mypara{C3: Processing latency estimation for dynamic workloads.}
Even with network latency estimation, the edge scheduler must also anticipate how long requests will take to process. 
Request processing times vary with workloads and resource contention, and are hard to predict without intrusive application changes. 
This makes lightweight yet accurate estimation essential for practical deployment.

\subsection{Key Ideas}
\label{sec:insights}

Our core insight is that standard 5G protocols and MEC application behaviors already expose the necessary signals for decoupled SLO-aware scheduling.
\sys{} exploits these readily available signals with no or minimal application modifications.
This approach enables three key ideas that address the fundamental challenges of decoupled scheduling:

\mypara{I1: Exploiting 5G control signal patterns for request identification (\autoref{sec:request-identify}).}
Standard 5G control signaling between UE and base station naturally exhibits distinctive patterns when new application requests are generated.
We find that buffer status reports (BSRs) and scheduling requests (SRs) provide reliable signatures that correlate with when the client sends a new request.
\sys{} leverages these existing control signals to detect when new requests begin at the RAN without payload inspection or protocol modifications.

\mypara{I2: Leveraging downlink stability for network latency estimation (\autoref{sec:reference-time}).}
5G downlink transmission characteristics provide inherent signals for network latency estimation.
We observe that downlink transmissions exhibit more predictable latency than uplink transmissions due to more wireless slots allocated for downlink, stable base station transmission power, and absence of scheduling jitter.
\sys{} exploits this asymmetry in 5G protocol behavior through lightweight probing that exchanges small timing packets between edge servers and client devices, enabling accurate latency tracking without operator infrastructure coordination.

\mypara{I3: Utilizing application lifecycle events for processing time prediction (\autoref{sec:proactive-resource-allocation}).}
MEC applications' request-response behaviors expose key lifecycle events that enable processing time estimation.
\sys{} tracks these naturally occurring events through server-side APIs and builds execution history without requiring invasive application changes.
We show this lightweight approach provides sufficient accuracy for deadline-aware scheduling while maintaining practicality.

\subsection{System Architecture}
\label{sec:architecture}

\begin{figure}[]
    \centering
    \includegraphics[width=1\columnwidth]{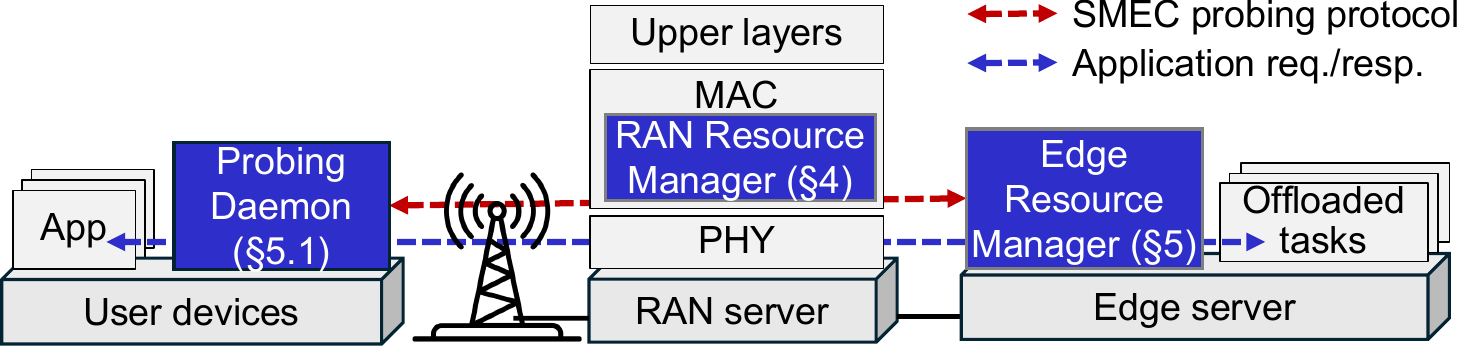}
    \tightcaption{\sys{} architecture: Blue boxes represent the main components of \sys{}, while grey boxes denote existing elements in the MEC stack. 
    For brevity, the 5G packet core and upper 5G protocol layers are omitted.}
    \label{fig:architecture}
\end{figure}

\autoref{fig:architecture} illustrates the \sys{} architecture, which consists of two main components that operate independently:

\mypara{RAN resource manager (\autoref{sec:ran_sched})} operates at the MAC layer and monitors UE--RAN control signal patterns to identify application request boundaries.
It dynamically allocates RAN resources based on each request's remaining time budget, increasing resource allocations when requests approach their deadlines to accelerate transmission.

\mypara{Edge resource manager (\autoref{sec:edge_sched})} estimates network transmission delays using a timing protocol between the UE-side daemon and server-side module.
It also estimates processing delays for incoming requests based on recent execution history to calculate remaining time budgets.
The manager then allocates heterogeneous compute resources (CPU and GPU) according to these time budgets and implements early drop mechanisms for requests unlikely to meet their deadlines.

\modified{\mypara{Specifying SLOs to the RAN.}
LC applications communicate their SLO requirements to the RAN through standard 5G interfaces without requiring custom protocols.
Edge servers can signal requirements through the Network Exposure Function (NEF) interface, or UE devices can convey QoS information when establishing Packet Data Unit (PDU) sessions~\cite{esti-5g-arch}.
\sys{} leverages 5G QoS Identifier (5QI) classes to map application SLOs, aligning with how commercial network operators classify traffic rather than requiring fine-grained per-application specifications.
This standards-compliant approach enables deployment within existing commercial MEC ecosystems while providing the necessary SLO information for effective resource management.}

%% file: 04_ran_sched.tex
\section{RAN Resource Management}
\label{sec:ran_sched}

This section presents the design of our SLO-aware RAN resource management, which operates at the RAN's MAC layer that makes a decision on RAN resource allocation.

\subsection{Identifying Application Requests}
\label{sec:request-identify}

\begin{figure}[t]
    \centering
    \includegraphics[width=\columnwidth]{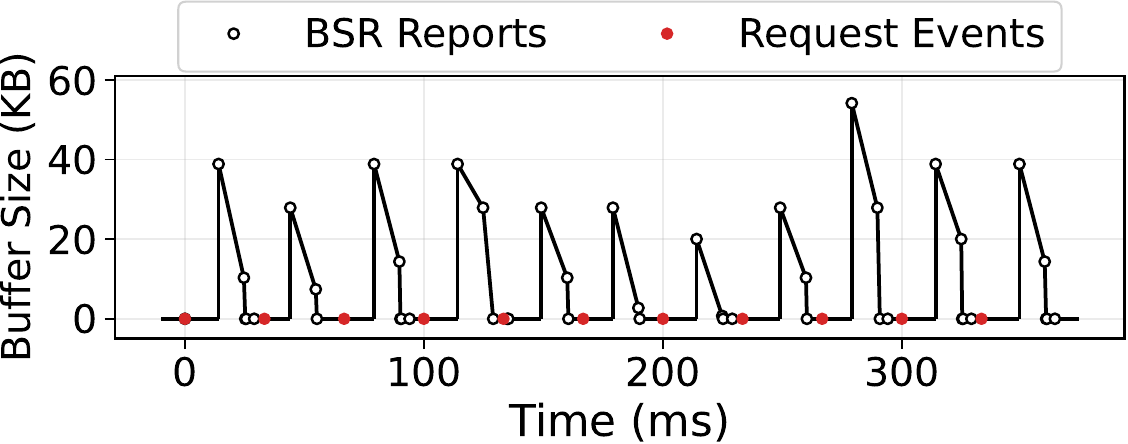}
    \tightcaption{Correlation between bytes reported in BSR and application requests.}
    \label{fig:bsr-correlation}
\end{figure}

As described in~\autoref{sec:challenges}, RAN protocol layers lack visibility into packet payloads, preventing direct identification of application request boundaries.
Even if payload access were available (\eg once the I/Q samples are demodulated into user data bits at the Packet Data Convergence Protocol (PDCP) layer), parsing application-layer data would violate the strict timing requirements of MAC scheduling, which must complete decisions within \us{500} or \ms{1} to meet 5G timing requirement.

\modified{\mypara{Our approach.}
Our key observation is that control signals from the UE to the RAN's MAC layer, particularly Buffer Status Reports (BSR), exhibit strong correlation with application-level requests.
As shown in \autoref{fig:bsr-correlation}, when a UE generates a new application request, additional data enters its uplink buffer, causing a noticeable increase in the reported BSR value.
This provides a MAC-layer signal that reveals request boundaries without requiring payload inspection.

We leverage this correlation to detect application request boundaries through BSR increases.
Specifically, the MAC layer identifies a new request boundary when the UE's reported BSR exhibits a step increase compared to the previous report.
We define the request start time, $t_{start}$, as the timestamp when the MAC scheduler receives the first BSR report reflecting this increase.

This approach achieves per-request granularity when requests arrive with inter-generation times exceeding the BSR update interval.
Under high sending rates or bursty workloads, multiple requests may be generated within a single BSR interval, appearing as one aggregated increase in the BSR.
In such cases, the MAC layer treats these as a single \emph{request group} sharing one $t_{start}$, and our scheduling decisions apply at the group level.

To handle multiple traffic types on the same UE (\eg traffic from different applications running concurrently), we leverage 5G logical channel groups (LCGs), which allow UEs to report separate BSRs per LCG to the MAC scheduler.
By configuring uplink traffic into different LCGs based on SLO classes, the scheduler can track buffer status and infer $t_{start}$ per traffic class rather than per UE.

Note that control signaling operates independently of user data transmission.
The 5G protocol assigns higher priority to BSR transmission than to user data, ensuring that the MAC scheduler can observe buffer state changes even under heavy traffic conditions.}

\subsection{Deadline-aware RAN Scheduling}
\label{sec:mac-scheduling}

Once the MAC layer identifies the start time of a new request, it computes the remaining time budget as:
\begin{equation}
    t^{RAN}_{budget} = \text{SLO} - (t_{current} - t_{start})
\end{equation}

With this information, the scheduler needs to decide how to allocate wireless resources across competing requests.

A 5G MAC layer allocates wireless resources\footnote{Physical Resource Blocks in 5G terminology} based on conventional metrics such as proportional fairness, which considers factors like channel quality and historical average throughput to balance efficiency and fairness among UEs.
However, these approaches do not account for application SLO requirements.

Unlike prior designs~\cite{xu2022tutti,arma:mobisys25} that emphasize fairness between LC and BE requests, our scheduler explicitly prioritizes LC requests based on their remaining time budgets.
This design choice stems from a key observation: the subsequent compute stage at the edge server introduces additional latency that the RAN cannot accurately observe or account for due to limited visibility and coordination across operators.
Therefore, our scheduler aims to minimize deadline violations for LC requests while ensuring starvation freedom for BE requests when resources remain available.

To achieve this, we allocate resources to LC requests as quickly as possible, preserving sufficient time budget for the compute stage to meet end-to-end deadlines.
Among competing LC requests, our scheduler prioritizes requests with the smallest remaining time budget.
This ensures that requests approaching their deadlines receive higher priority, with already-violated requests receiving maximum priority to prevent buffer blocking.

\mypara{Ensuring starvation freedom for BE requests.}
While prioritizing LC requests, our design ensures that BE requests remain starvation-free through two mechanisms.
\modified{First, we assign higher priority to Scheduling Request (SR)-triggered resource allocations.
In standard 5G MAC scheduling, when a UE has not received resources for an extended period, it sends an SR, a control signal that requests uplink grants from the MAC scheduler.
Our scheduler assigns these SR-triggered allocations higher priority than regular scheduling decisions (even higher than LC requests), ensuring that BE UEs maintain forward progress.
Since SR-triggered allocations are small (typically 1--2\% of the wireless resources available for a slot), they do not impact LC request performance.
Second, we implement dynamic priority reset: when an LC request completes transmission (detected when its BSR reaches zero), we immediately reset the UE's priority to zero, allowing BE UEs to fully utilize available resources and ensuring efficient bandwidth utilization.}

%% file: 05_edge_sched.tex
\section{Edge Resource Management}
\label{sec:edge_sched}

While the RAN resource manager prioritizes LC requests during uplink transmission, the edge server must allocate compute resources to meet SLO deadlines without any coordination with the RAN.
The key challenge is estimating each request's remaining time budget by determining the network latency already consumed (uplink), the future network latency (downlink), and the expected processing time.
\modified{This section presents our approach that leverages timing signals naturally generated by applications during their request-response lifecycle, which can be captured via a lightweight API (\autoref{tbl:smec-api}).}

\begin{table}[]
\setlength{\tabcolsep}{10pt}
\renewcommand{\arraystretch}{1.0}
\centering
\footnotesize
\begin{tabular}{@{}ll@{}}
\toprule
\textbf{API Call} & \textbf{Purpose} \\ \midrule
\texttt{request\_sent(req\_data*)} & Report new request sent \\
\texttt{request\_arrived(req\_data*)} & Report new request arrival \\
\texttt{processing\_started(req\_id)} & Report processing start \\
\texttt{processing\_ended(req\_id)} & Report processing completion \\
\texttt{response\_sent(resp\_data*)} & Report response transmission \\ 
\texttt{response\_arrived(resp\_data*)} & Report response arrival \\\bottomrule
\end{tabular}
\vspace{-0.5em}
\caption{\sys{} API for network and processing time estimation.}
\label{tbl:smec-api}
\end{table}

\subsection{Estimating Network Latency}
\label{sec:reference-time}

Consider a request with a \ms{100} SLO that has spent \ms{60} in uplink transmission. 
The edge server has only \ms{40} remaining to complete processing and downlink transmission.
Without accurate timing information about both consumed uplink delays and future downlink delays, the edge server cannot distinguish between urgent requests approaching their deadlines and requests with ample time remaining.

However, measuring per-request network latency is challenging: the edge server only observes when requests arrive locally but lacks visibility into when they were originally sent from the client, the uplink delays they experienced, or the downlink delays that responses will experience.

\mypara{Possible approach.}
A straightforward approach would piggyback a sending timestamp in each request, enabling the server to compute transmission delay upon reception.
Unfortunately, this requires precise time synchronization between the UE and server, which is infeasible in MEC environments.
Network Time Protocol (NTP)~\cite{ntp-rfc5905} introduces synchronization errors ranging from tens to hundreds of milliseconds, which are larger than the time budget of LC applications.
PTP~\cite{ptp-std} is also unsuitable because it assumes network delays are symmetric.
5G networks exhibit inherent asymmetry where uplink latency is both higher and more variable than downlink latency due to protocol design and UE energy constraints. 

\begin{figure}[]
    \centering
    \includegraphics[width=0.4\textwidth]{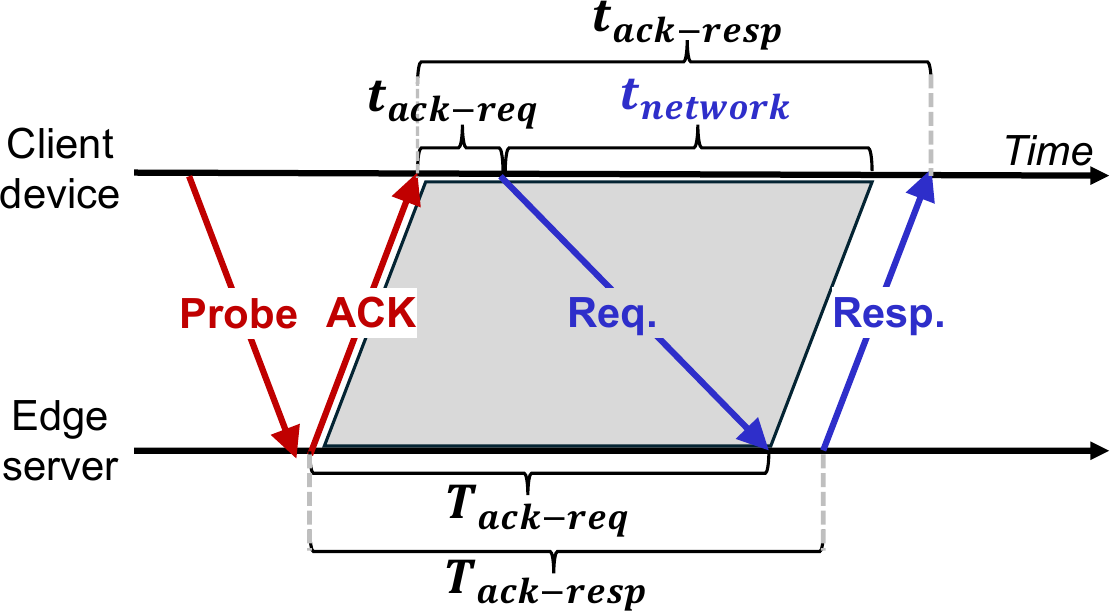}
    \vspace{-0.5em}
    \caption{Probing-based network latency estimation. 
            Red arrows indicate probing protocol packets while blue arrows indicate application requests and responses.}
    \label{fig:reference-time}
    \vspace{-1em}
\end{figure}

\mypara{Our approach.}
We leverage our observation that while uplink latency in 5G networks is highly variable, downlink latency remains consistently stable (\autoref{fig:motive-decompose}).
We exploit the stability of downlink through a lightweight \emph{probing-based network latency estimation} protocol that establishes timing references using the stable downlink path.

As illustrated in \autoref{fig:reference-time}, the client device periodically sends probe packets to the server, which responds with ACKs over the stable downlink.
To implement this protocol, \sys{} runs a per-UE \emph{probing daemon} on the client side (shown in~\autoref{fig:architecture}) and integrates the corresponding functionality into the edge resource manager on the server side.
When an application sends a request, it reports the request to the timing daemon via the \texttt{request\_sent} API (\autoref{tbl:smec-api}), which measures $t_{ack-req}$ (the time elapsed since receiving the latest ACK) and inserts this timing metadata into the request payload.
Upon receiving the request, the server computes $T_{ack-req}$ (the time difference between sending the most recent ACK and receiving the current request).
The stable downlink timing creates a \emph{parallelogram} relationship (grey region in \autoref{fig:reference-time}), allowing the server to estimate network latency as $T_{ack-req} - t_{ack-req}$.

However, if the response size is significantly larger than the ACK size, there will be a gap in downlink transmission latency between the two (\autoref{fig:motive-decompose}). 
To compensate for this difference, when the server sends a response, it computes the elapsed time since it sent the last ACK ($T_{ack-resp}$) and reports this value to the client daemon as part of the response. 
Upon receiving the response, the client daemon uses this information to compute the time elapsed since it received the last ACK ($t_{ack-resp}$) and calculates a compensation factor ($t_{comp}$), which it reports to the server as part of the next probe. 
The client daemon maintains this compensation factor separately for each application. 
Using this correction factor, the server estimates the network latency while accounting for response size differences:
\begin{equation}
    t_{network} = T_{ack-req} - t_{ack-req} + t_{comp}
\end{equation}

To handle packet losses, each probe-ACK exchange carries a unique ID that both endpoints use to synchronize on the most recent successful exchange.
\modified{The design incurs minimal overhead: the client sends small probe/ACK packets (\textless{}\byte{100}) every few seconds, but \emph{only while the UE is actively serving LC traffic}.
When the UE is idle, the probing daemon pauses, avoiding interference with the UE's power-saving mechanism (\eg Discontinuous Reception (DRX)).}

This approach extends to distributed scenarios where request initiators and response receivers are different devices (\eg smart stadium with camera initiators and audience UE receivers).
In such cases, the initiator sends probe packets for network latency estimation while receivers sends probe packets to report compensation factors, enabling accurate end-to-end latency estimation across the distributed path.

\subsection{Estimating Remaining Time Budget}
\label{sec:deadline-estimation}

Given the network latency estimates, the resource manager now needs to predict each request's \emph{processing time} to compute an accurate remaining time budget.
However, accurately predicting processing time while maintaining practical deployability presents challenges.
Applications exhibit variable processing times due to workload change and resource contention, making accurate prediction difficult.

\mypara{Possible approaches.}
One approach would be to infer processing time from system-wide signals like CPU utilization, scheduling frequency, and hardware performance counters (\eg cache misses, memory footprint), but such methods suffer from either high inference overhead or poor accuracy, as shown by existing work~\cite{liu2023intelligent, peng2024ufo}.
Alternatively, one could extensively instrument applications to measure real-time workload information (\eg Caladan~\cite{caladan:osdi20}); while this obtains precise timing information, it requires significant application modifications, limiting practicality.

\mypara{Our approach.}
Building on the client-side request tracking for network latency estimation (\autoref{sec:reference-time}), we extend this to the server side where applications report key lifecycle events (\eg request arrival, processing start/end) to the resource manager via the \sys{} API (\autoref{tbl:smec-api}).
By tracking these events, the resource manager estimates processing delays without requiring detailed application knowledge or extensive instrumentation.

The resource manager tracks two key metrics: waiting time ($t_{wait}$) defined as the time elapsed from request arrival until processing begins, and processing time using the median of last $R$ requests as a robust predictor ($t_{process}$).
While this median-based approach is simple and may introduce some prediction error, it performs well in practice (\autoref{sec:eval-micro-benchmarks-network-and-processing-time-estimation}) while minimizing application modifications.
The manager can then compute the remaining time budget as:
\begin{equation}
    t^{edge}_{budget} = SLO - (t_{network} + t_{wait} + t_{process})
\end{equation}

\subsection{Deadline-aware Proactive Edge Resource Scheduling}
\label{sec:proactive-resource-allocation}

\begin{figure}[]
    \centering
    \begin{subfigure}[b]{0.48\columnwidth}
        \centering
        \includegraphics[width=\textwidth]{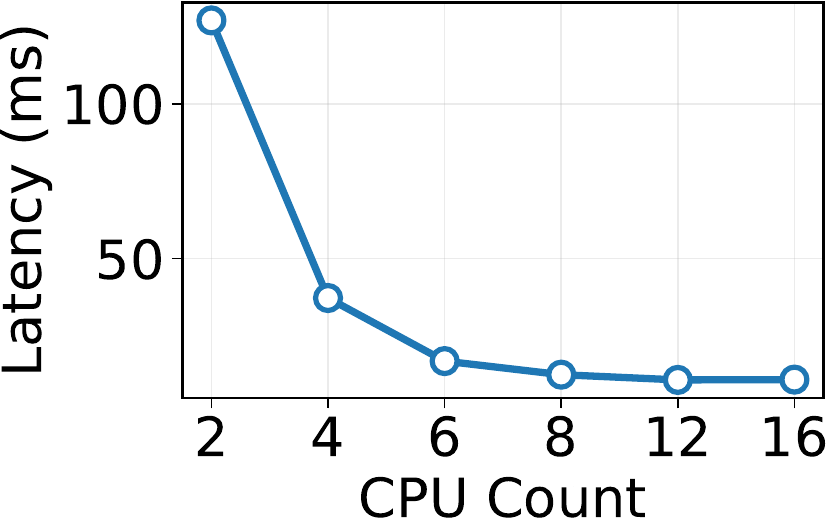}
        \caption{CPU-based task}
        \label{fig:latency-cpu}
    \end{subfigure}
    \hfill
    \begin{subfigure}[b]{0.48\columnwidth}
        \centering
        \includegraphics[width=\textwidth]{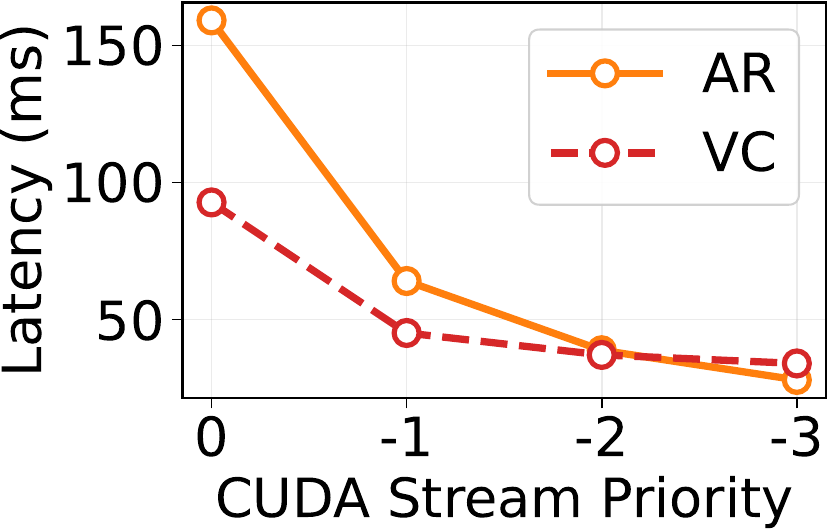}
        \caption{GPU-based task}
        \label{fig:latency-gpu}
    \end{subfigure}
    \vspace{1em}
    \tightcaption{Relationship between compute resources allocation and process latency.}
    \label{fig:latency-resource}
\end{figure}

\modified{The resource manager determines urgency based on each request's remaining time budget.
We define an urgency threshold as a fraction $\tau$ of the application's SLO (default $\tau=0.1$), marking a request as urgent when $t^{edge}_{budget} < \tau \times SLO$.
When a request becomes urgent, the manager \emph{proactively} allocates additional resources to prevent SLO violations.

\sys{} focuses on CPU and GPU management, which constitute the primary compute bottlenecks in edge servers.
CPU cores are heavily contended among multiple applications, while GPUs accelerate ML inference and video processing.
\autoref{alg:edge-proactive-sched} in \autoref{appnd:edge-scheduling-algo} summarizes our deadline-aware scheduling approach.}

\mypara{CPU management.}
Our CPU manager partitions cores across applications using CPU affinity to ensure isolation and predictable performance.
Based on the observation that increasing CPU core allocation reduces processing latency (\autoref{fig:latency-cpu}), the manager allocates additional cores to urgent requests when they risk missing deadlines.
This policy is most effective when an application can parallelize request processing (\eg via multi-threaded execution).
To avoid thrashing from frequent core reallocations, the manager enforces a brief cool-down period (\eg \ms{100}) after each allocation, and assigns another core only if requests still risk missing deadlines after the cool-down.
This mechanism prevents wasteful oscillations while ensuring urgent requests receive the resources they need.

For \emph{reclamation}, we use average CPU utilization rather than urgency signals.
Urgency-based reclamation creates instability: removing a single core from a latency-critical application can cause an abrupt shift from meeting deadlines to missing many, leading to scheduler thrashing.
Instead, when an application's utilization falls below a threshold (\eg 60\%), the manager safely reclaims cores, providing stable resource management without oscillatory behavior.

\mypara{GPU management.}
Commercial MEC offerings~\cite{www-outpost,www-azure-stack-edge} typically deploy inference-optimized GPUs such as NVIDIA L4~\cite{nvidia-l4} and T4~\cite{nvidia-t4}.
These devices lack hardware-level partitioning capabilities like Multi-Instance GPU (MIG)~\cite{nvidia-mig}.
\modified{Even when hardware partitioning is available, it only supports static allocation at application launch and cannot be dynamically adjusted during runtime.

To enable dynamic GPU scheduling, we leverage CUDA stream priorities, inspired by Orion~\cite{strati2024orion}.
NVIDIA's Multi-Process Service (MPS)~\cite{nvidia-mps} enables multiple applications to share the GPU while maintaining a unified priority hierarchy across their CUDA streams.
When kernels from multiple applications contend for GPU resources, those launched from higher-priority streams receive preferential scheduling.
This mechanism allows the resource manager to influence GPU scheduling decisions without requiring hardware partitioning or GPU driver modifications.

We exploit this capability by assigning stream priorities at request granularity based on urgency.
As shown in \autoref{fig:latency-gpu}, higher stream priorities reduce processing latency under contention.
The resource manager assigns higher-priority streams to requests whose expected processing time closely matches their remaining time budget, while requests with slack use lower-priority tiers.
This ensures urgent requests receive preferential GPU access without starving other workloads.

\mypara{Early drop: handling overly urgent requests.}
When a request becomes overly urgent (\eg $t^{edge}_{budget}\leq 0$) due to excessive network or queuing delays, no amount of compute resources can recover the already-elapsed time.
Processing such requests wastes resources that could serve other requests still capable of meeting their deadlines.
Therefore, when the edge server operates under load, the resource manager immediately drops overly urgent requests (``early drop''), redirecting resources to requests with viable time budgets.}

%% file: 06_impl.tex
\section{Implementation}
\label{sec:impl}

We have implemented a prototype of \sys{} consisting of the RAN resource manager and edge resource manager, comprising approximately 7,700 lines of Python and C++ code~\cite{smec-github}.

\mypara{RAN resource manager.}
We implement the RAN resource manager in the srsRAN 5G's MAC layer~\cite{srsran} with our request identification mechanism and deadline-aware scheduling algorithm.
Since our design is not tightly coupled with srsRAN, it can be ported to other RAN stacks such as OAI 5G~\cite{oai5g:www}.

\mypara{Edge resource manager.}
We implement the edge resource manager as a user-space daemon that coordinates with applications through the \sys{} API.
It consists of four core components: network latency estimation, processing time prediction, CPU/GPU management modules described below.

\mypara{Network latency estimation.}
The network latency estimation module operates on both client and server to enable accurate per-request network latency measurement.
The client periodically sends a probe message containing a 4-byte compensation factor tagged with a 4-byte probe ID to the server.
The server replies with 12-byte ACK packets containing the same probe ID and the ACK's sending timestamp over the stable downlink path.
In our prototype, we use a 1-second probing frequency to balance accuracy with overhead.

\mypara{Processing time prediction.}
The processing estimation module maintains a sliding window of the past $R$ requests' processing times for each application to predict future delays.
By tracking both queueing and processing components through the \sys{} API events, the system builds application-specific performance profiles.
We use $R=10$ as the window size in our prototype, providing sufficient history while remaining responsive to workload changes.

\mypara{CPU management.}
Our CPU manager leverages Linux's \texttt{sched\_setaffinity} system call~\cite{sched-setaffinity:manpage} to dynamically bind application processes to specific CPU cores based on deadline urgency.
This user-space approach enables fine-grained CPU allocation and reclamation without requiring kernel modifications or custom scheduling classes. 

\mypara{GPU management.}
Our GPU scheduler operates through NVIDIA's MPS, which enables CUDA stream priorities from different application processes to be compared on a unified scale.
For example, a stream with priority -3 from one process correctly receives higher priority than a stream with priority 0 from another process.
Each application creates multiple CUDA streams at initialization using the \texttt{cudaStreamCreateWithPriority()} API~\cite{cuda-stream-priority}, with each stream assigned a distinct priority level.
Based on deadline urgency feedback from the resource manager, incoming requests are dispatched to the appropriate stream, ensuring urgent requests execute on high-priority streams while less critical requests use lower-priority streams.

%% file: 07_eval.tex
\section{Evaluation}
\label{sec:eval}

We evaluate \sys{} on our private 5G MEC testbed that emulates commercial deployments with real applications. 

\subsection{Experimental Setup}
\label{sec:eval-setup}

\mypara{Testbed setup.} 
Our 5G MEC testbed consists of a UE emulator (Amari UE Simbox~\cite{amari-uesim}), two x86 servers, and a USRP X310~\cite{usrp-x310} radio unit.
\textbf{Server-1} acts as the RAN, running srsRAN~\cite{srsran} and Open5GS~\cite{open5gc} with Intel Xeon Silver 4310 CPU, 128 GB memory, and Ubuntu 22.04.
We configure the RAN in TDD mode with 80 MHz bandwidth and 2$\times$2 MIMO on band 78, representing typical 5G deployments.
\textbf{Server-2} serves as the edge server with NVIDIA L4 GPU~\cite{nvidia-l4}, Intel Xeon Gold 6430 CPU (32 cores), 256 GB memory, and Ubuntu 24.04, reflecting commercial MEC offerings~\cite{www-outpost}.
To emulate CPU contention, we disable hyper-threading and use 24 cores.
Servers are connected via 25 GbE.

\mypara{Applications.}
We implement three LC applications (\autoref{tbl:apps}) and one BE application. 
All LC applications are video-based, and we treat each video frame as a single request.

\noindent\textit{\textbf{Smart stadium (SS)}} (SLO: \ms{100})~\cite{www-orange-5gstadium,www-red5perfect5G-in-venue}: 5G-enabled cameras upload high-resolution 4K video streams to the edge server, which transcodes each stream into multiple lower-bit-rate versions and delivers them to subscribing clients.
In our evaluation, we use the same UE to emulate both the camera and subscribing clients.
This represents a CPU-intensive workload for live streaming services.
We stream videos over RTP and implement transcoding using FFmpeg's H.264 codec.
We use a video from the AdaPool dataset~\cite{stergiou2021adapool}, re-encoded as a 4K 60 fps stream at \Mbps{20}.

\noindent\textit{\textbf{Augmented reality (AR)}} (SLO: \ms{100})~\cite{mao2019delay,padmanabhan2023gemel}: AR devices stream videos over RTP to the edge server, which performs object detection using a YOLO model~\cite{yolo:cvpr16} and sends annotated results back to the AR devices.
This represents a GPU-intensive workload for computer vision applications.
We use a video from the MOT dataset~\cite{leal2015motchallenge}, re-encoded as a 1080p 30 fps stream at \Mbps{8}.

\noindent\textit{\textbf{Video conferencing (VC)}} (SLO: \ms{150})~\cite{www-zoom-latency,www-wikipedia-audio-latency}: Client devices with limited connectivity send low-quality video streams to the edge server, which enhances them using super-resolution and streams the enhanced video back to the clients.
This represents a GPU-intensive video enhancement workload.
We stream videos over RTP and implement super-resolution using the Real-ESRGAN model~\cite{wang2021realesrgan}.
We use a video from the ICME-VSR dataset~\cite{naderi2025icme}, re-encoded as a 320p 30 fps stream at \Kbps{800}.

\noindent\textit{\textbf{File transfer (FT)}} (No SLO): Client devices transfer files with dummy content to a remote server (not the edge server) to simulate best-effort traffic.

\mypara{Application workloads.}
To evaluate the impact of workload characteristics on \sys{}, we use two types of workloads:

\noindent \textit{\textbf{Static}}:
To evaluate how \sys{} performs under sustained heavy load on both compute resources and the RAN, we design a static workload that creates continuous pressure on the system.
We use 12 concurrent UEs: 2 for smart stadium, 2 for augmented reality, 2 for video conferencing, and 6 for file transfer.
The LC UEs continuously send video frames at their respective target rates.
The transcoding task of SS converts videos into three fixed resolutions (2K, 1080p, 720p), while AR performs object detection using the YOLOv8 medium model~\cite{yolov8}.
FT keeps sending \mbyte{3} files repeatedly.

\modified{\noindent \textit{\textbf{Dynamic}}: 
To evaluate how \sys{} handles bursty requests and resource contention, we design a dynamic workload with fluctuating demand.
We use 2 UEs for SS, but the transcoding task randomly varies the number of target resolutions (between 2 and 4), creating fluctuating compute demand.
For AR and VC, we vary the number of UEs sending requests dynamically between 0 and 2.}
To amplify computational bursts, AR uses the larger YOLOv8 large model~\cite{yolov8}.
We use 6 UEs for FT, and each UE repeatedly uploads files with sizes uniformly chosen between \kbyte{1} and \mbyte{10}.

\mypara{Baselines.} 
We compare the performance of \sys{} against three baselines: (1) Default scheduler (\textbf{Default}), (2) \textbf{\tutti{}}, and (3) \textbf{\arma{}}.
For the default scheduler, the RAN uses the PF scheduler, while the edge server employs the default Linux scheduler (EEVDF~\cite{eevdf}) for CPU processes and the hardware scheduler in the L4 GPU for GPU tasks. 
For \tutti{} and \arma{}, since neither considers edge resource scheduling, we pair them with the default scheduler at the edge server.
\modified{To ensure a fair comparison, we implement early drop (\autoref{sec:proactive-resource-allocation}) at the edge server for all baselines based on application queue length: we set the queue length to 10 and drop incoming requests when the queue exceeds this threshold.}

\subsection{Performance under Static Workloads}
\label{sec:eval-e2e-static}

\begin{figure}[]
    \centering
    \includegraphics[width=\columnwidth]{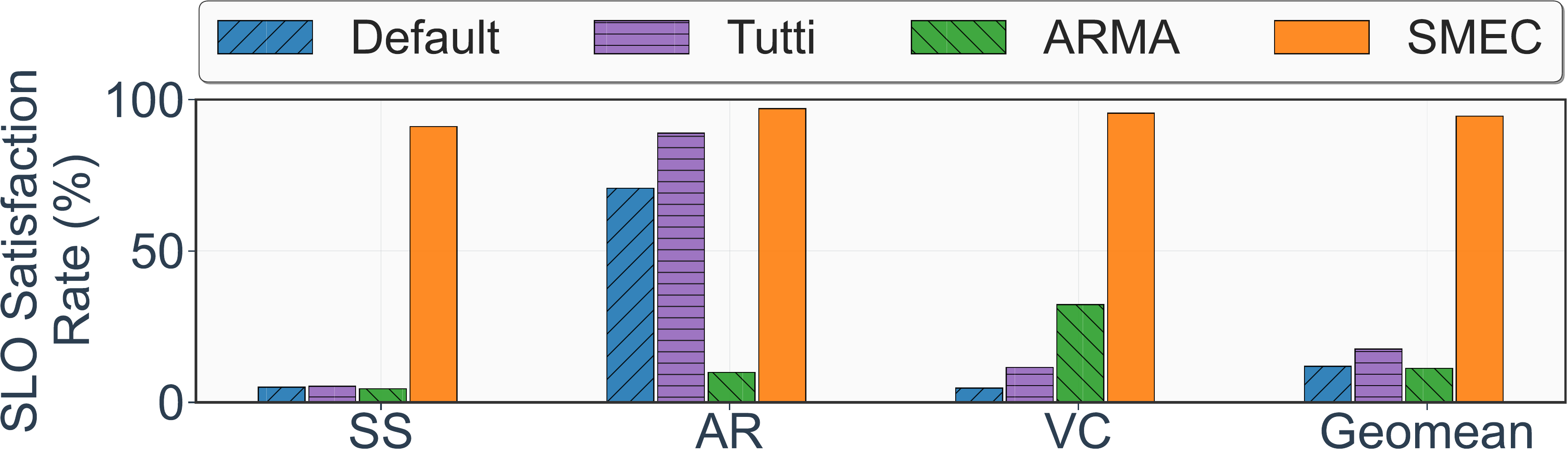}
    \tightcaption{SLO satisfaction rate under static workload.}
    \label{fig:slo-statistics-steady-workload}
\end{figure}

\begin{figure}[]
    \centering
    \includegraphics[width=\columnwidth]{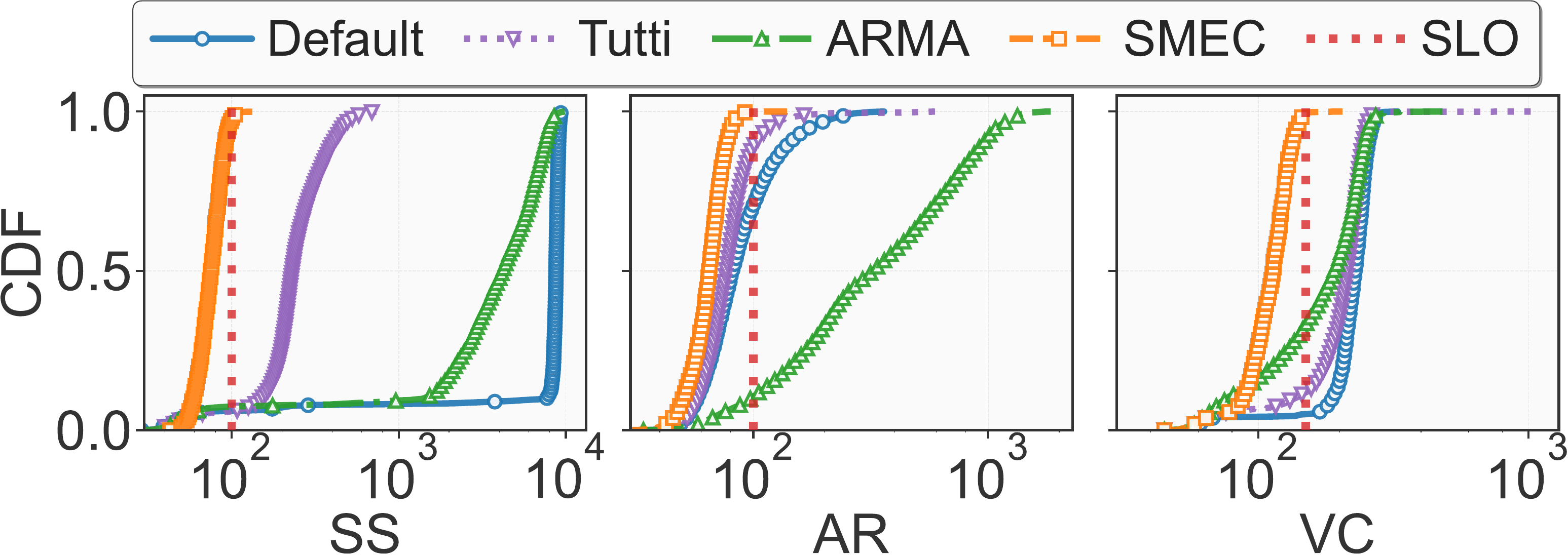}
    \tightcaption{End-to-end latency (ms) under static workload.}
    \label{fig:cdf-steady-workload}
\end{figure}

We first evaluate \sys{} under static workloads that create sustained pressure on both network and compute resources.

\mypara{SLO satisfaction.}
Across all three applications, \sys{} exceeds 90\% SLO satisfaction and outperforms the baselines (\autoref{fig:slo-statistics-steady-workload}).
For SS, \sys{} reaches 91\% versus $<$6\% for baselines; for VC, \sys{} maintains 96\% compared to only 5--35\% for baselines.
For AR, contention is modest under the static workload, so the headroom is smaller: \sys{} improves SLO satisfaction by about 8\% points over \tutti{} and about 26\% points over Default.

\mypara{Tail latency.}
\sys{} substantially reduces tail latency (\autoref{fig:cdf-steady-workload}): for smart stadium, P99 latency drops by 89$\times$, 5.6$\times$, and 84$\times$ relative to the Default, \tutti{}, and \arma{}, respectively.
For AR, \sys{} reduces P99 latency by 2.9$\times$, 2$\times$, and 15.6$\times$ relative to the Default, \tutti{}, and \arma{}, respectively.
For VC, \sys{} shows a smaller tail-latency gain ($\approx$2$\times$) because its uplink demand is low; VC is primarily impacted by compute contention rather than network latency.

\begin{figure}[]
    \centering
    \includegraphics[width=\columnwidth]{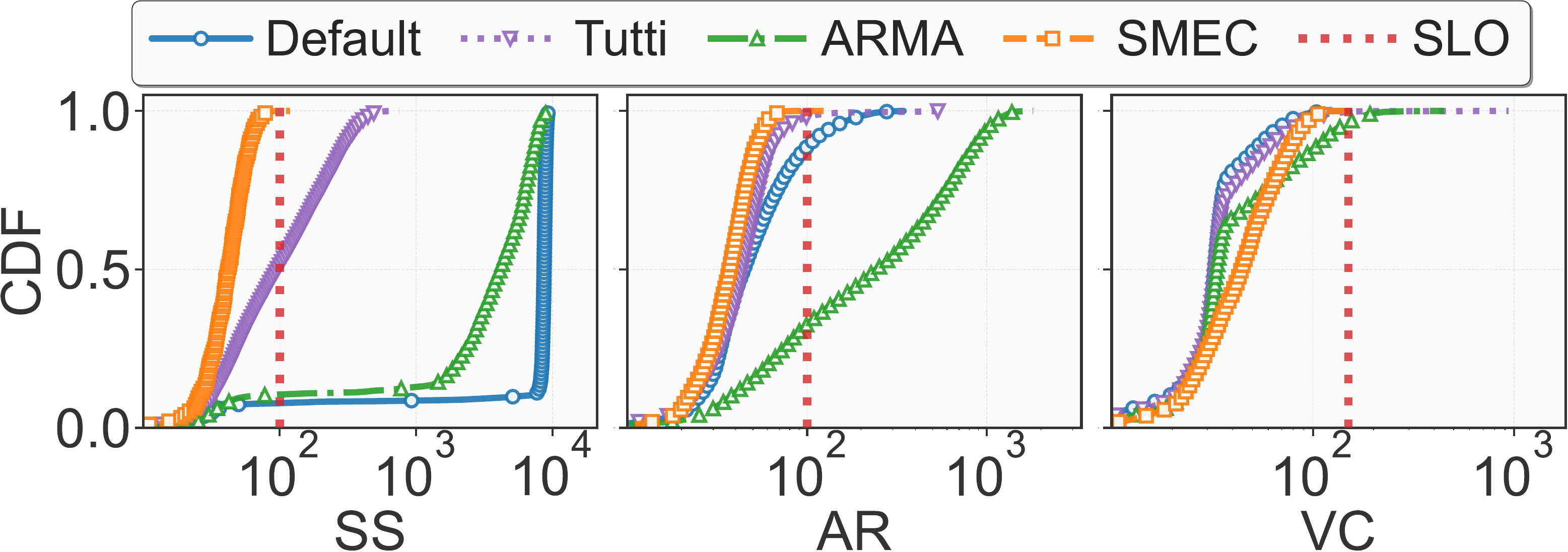}
    \tightcaption{Network latency (ms) under static workload.}
    \label{fig:network-latency-steady-workload}
\end{figure}

\begin{figure}[]
    \centering
    \includegraphics[width=\columnwidth]{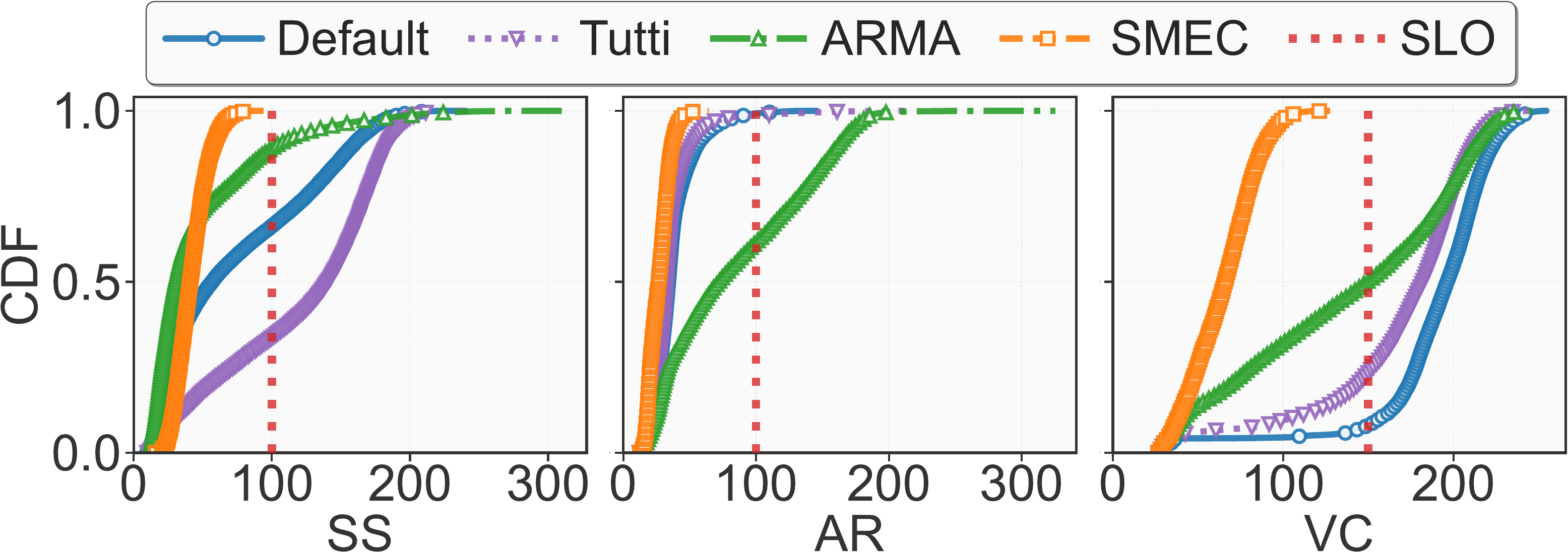}
    \tightcaption{Processing latency (ms) under static workload.}
    \label{fig:processing-latency-steady-workload}
\end{figure}

\mypara{Why baselines fall short.}
The latency breakdown pinpoints the root causes.
\textbf{\textit{Network}} (\autoref{fig:network-latency-steady-workload}).
Default and \arma{} rely on PF at the RAN, which allocates resources fairly across LC and BE UEs. 
This allows BE flows to occupy uplink resources that LC traffic urgently needs and starve LC apps. 
Uplink-heavy workloads suffer most: for SS, tail network latency approaches \second{10}, leading to about \percent{90} of requests missing their SLO. 
\tutti{} also causes over \percent{50} SLO violations at network side for SS because it depends on server-side notifications to infer request start times; this delay prevents timely acceleration of urgent requests, leading to SLO violations.
\textbf{\textit{Edge server}} (\autoref{fig:processing-latency-steady-workload}).
All baselines ignore the effect of compute contention. 
Consequently, \tutti{} sees $\sim$\percent{70} of SS requests misses SLO deadlines under CPU contention, and across Default, \tutti{}, and \arma{}, video conferencing experiences $\sim$50--\percent{90} SLO violations dominated by GPU contention.
Notably, for smart stadium, Default and \arma{} exhibit fewer processing-side SLO violations not because they handle compute contention, but because severe uplink congestion causes requests to backlog at the UE sending buffer.
When this buffer is full, some requests are dropped, which in turn lowers CPU load at the server.

\mypara{Why \arma{} performs much poorer for AR.}
Under resource pressure, \arma{}'s RAN scheduler and reallocates uplink resources away from AR to prioritize SS; as a result, AR receives fewer grants and its uplink waiting time increases. 
This raises AR's \textit{network} latency significantly (AR in \autoref{fig:network-latency-steady-workload}). 
A second-order effect further hurts \textit{compute}: as AR grants resume, many backlogged AR requests arrive at the server nearly simultaneously, creating a burst that inflates queueing and causes deadline misses (AR in \autoref{fig:processing-latency-steady-workload}).

\subsection{Performance under Dynamic Workloads}
\label{sec:eval-e2e-dynamic}

\begin{figure}[]
    \centering
    \includegraphics[width=\columnwidth]{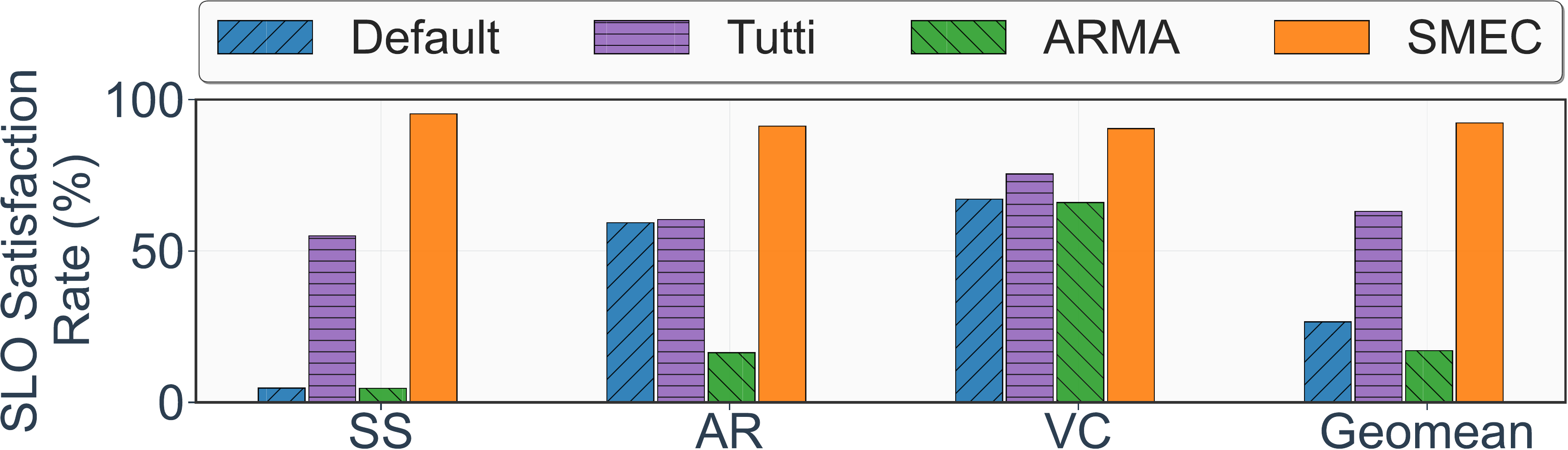}
    \tightcaption{SLO satisfaction rate under dynamic workload.}
    \label{fig:slo-statistics-dynamic-workload}
\end{figure}

\begin{figure}[]
    \centering
        \includegraphics[width=\columnwidth]{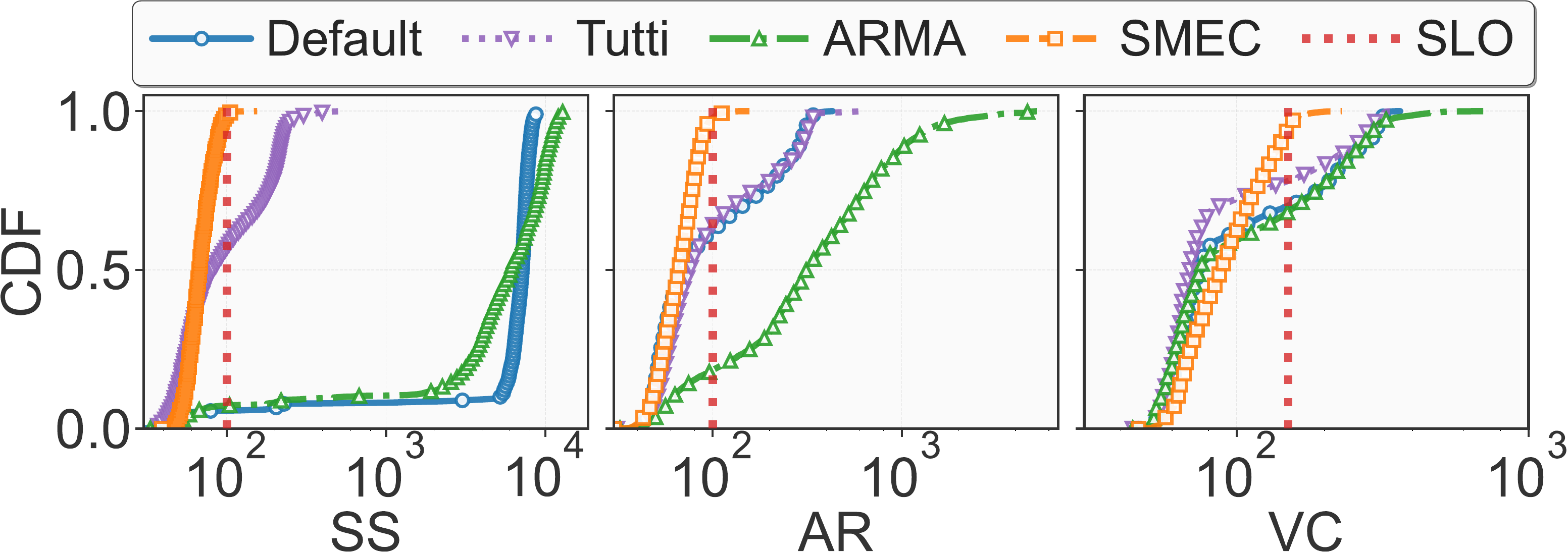}
        \tightcaption{E2E latency (ms) under dynamic workload.}
    \label{fig:cdf-dynamic-workload}
\end{figure}

We evaluate \sys{} under dynamic workloads that create bursty traffic patterns and fluctuating compute demands, emulating real-world edge environments with varying workloads.

\mypara{SLO satisfaction.}
\sys{} maintains over \percent{90} SLO satisfaction across all three LC applications under dynamic workloads (\autoref{fig:slo-statistics-dynamic-workload}), demonstrating robust performance even with fluctuating demands.
For SS, \sys{} achieves \percent{95.3} satisfaction compared to \percent{4.7} (Default), \percent{4.6} (\arma{}) and \percent{55} (\tutti{}), while for AR, \sys{} sustains \percent{91.2} \vs only \percent{59.3} (Default), \percent{16.4} (\arma{}) and \percent{60.4} (\tutti{}).
For VC, the gap is smaller (\percent{90.4} \vs 65--\percent{75} for baselines) because VC has low uplink demand and is affected mainly by GPU  contention rather than network latency. 
Under dynamic workloads, VC only misses deadlines during GPU bursts (\ie when all AR and VC clients issue requests concurrently) so SLO violations are concentrated in those burst windows, keeping the overall gap modest.

\mypara{Tail latency.}
\sys{} keeps tail latency within reasonable bounds, with delays rarely exceeding SLO targets by large margins (\autoref{fig:cdf-dynamic-workload}).
For SS, \sys{} reduces P99 latency by 87$\times$ compared to the default scheduler, while achieving 3.2$\times$ and 122$\times$ improvements over \tutti{} and \arma{}, respectively.
For AR, \sys{} reduces P99 latency by 3.2$\times$ \vs Default, 3.3$\times$ \vs \tutti{}, and 31$\times$ \vs \arma{}. 
For VC, P99 improves by $\sim$2$\times$; as in the static case, low uplink demand leaves VC compute-bound, so gains are smaller.

\begin{figure}[]
    \centering
    \includegraphics[width=\columnwidth]{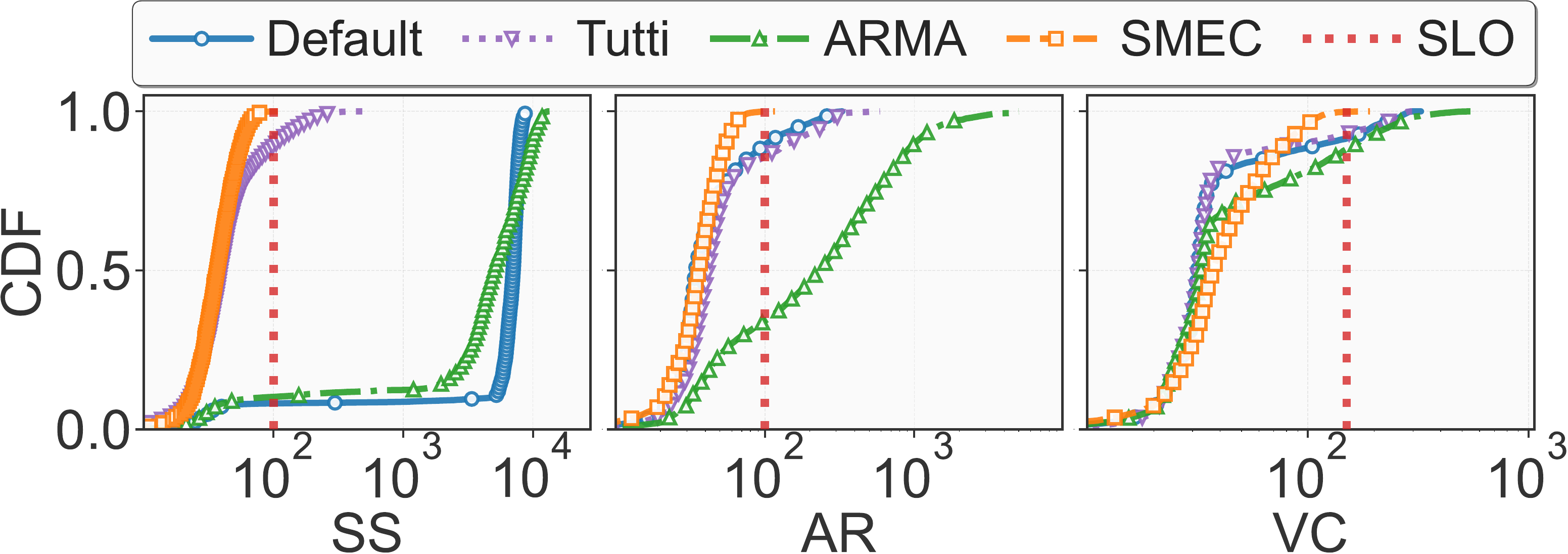}
    \vspace{-2em}
    \caption{Network latency (ms) under dynamic workload.}
    \label{fig:network-latency-dynamic-workload}
\end{figure}

\begin{figure}[]
    \centering
    \includegraphics[width=\columnwidth]{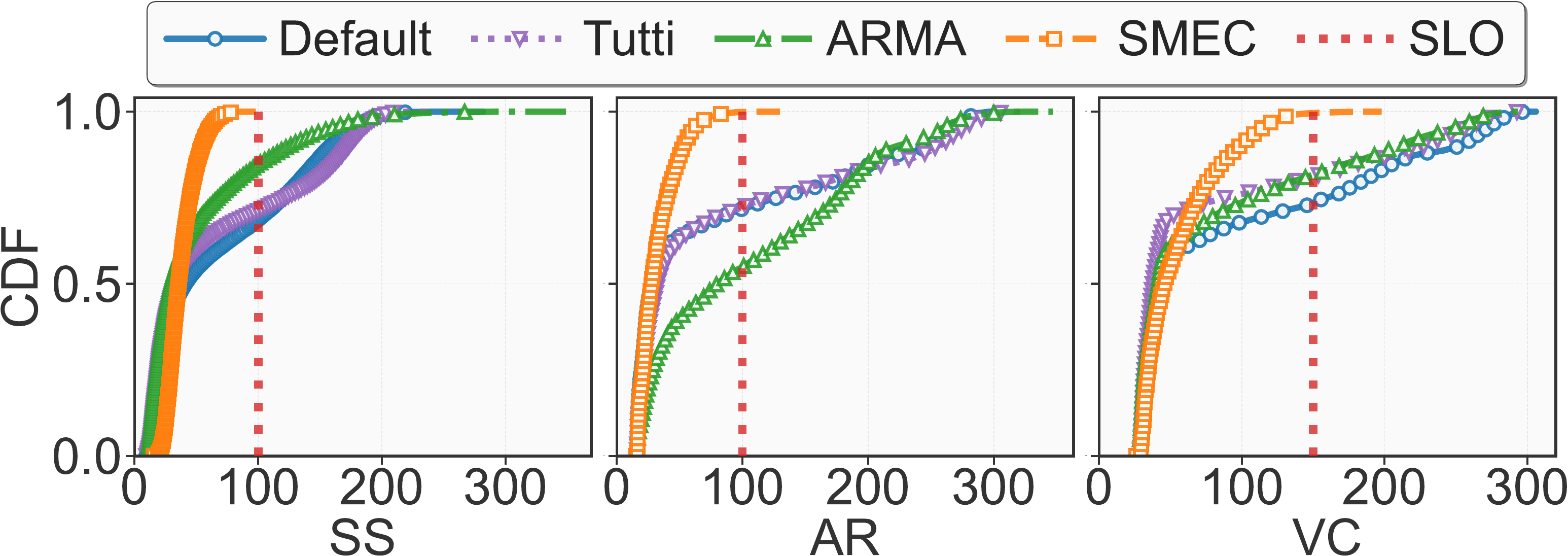}
    \tightcaption{Processing latency (ms) under dynamic workload.}
    \label{fig:processing-latency-dynamic-workload}
\end{figure}

\mypara{Why baselines fall short.}
To make the causes explicit, we similarly decompose end-to-end latency.
\textbf{\textit{Network}} (\autoref{fig:network-latency-dynamic-workload}). 
The reasons mirror the static case: Default and \arma{} use PF at the RAN, allowing BE flows to take uplink resources urgently needed by LC traffic, starving LC apps; \tutti{} depends on delayed server-side start inference, so it cannot accelerate urgent requests in time.
\textbf{\textit{Edge server}} (\autoref{fig:processing-latency-dynamic-workload}). 
The key difference in the dynamic setting is burstiness. 
Each burst overloads the edge server, creating long queues and around \percent{30} deadline misses for all baselines due to compute contention, which ignore compute contention. 
In contrast, \sys{} proactively controls backlog (\eg by dropping a small fraction of hopeless requests) to relieve pressure and keep queues short, thereby preventing burst-induced misses.

\subsection{Impact on Best Effort Applications}
\label{sec:eval-e2e-be}

\begin{figure}[]
    \centering
    \includegraphics[width=\columnwidth]{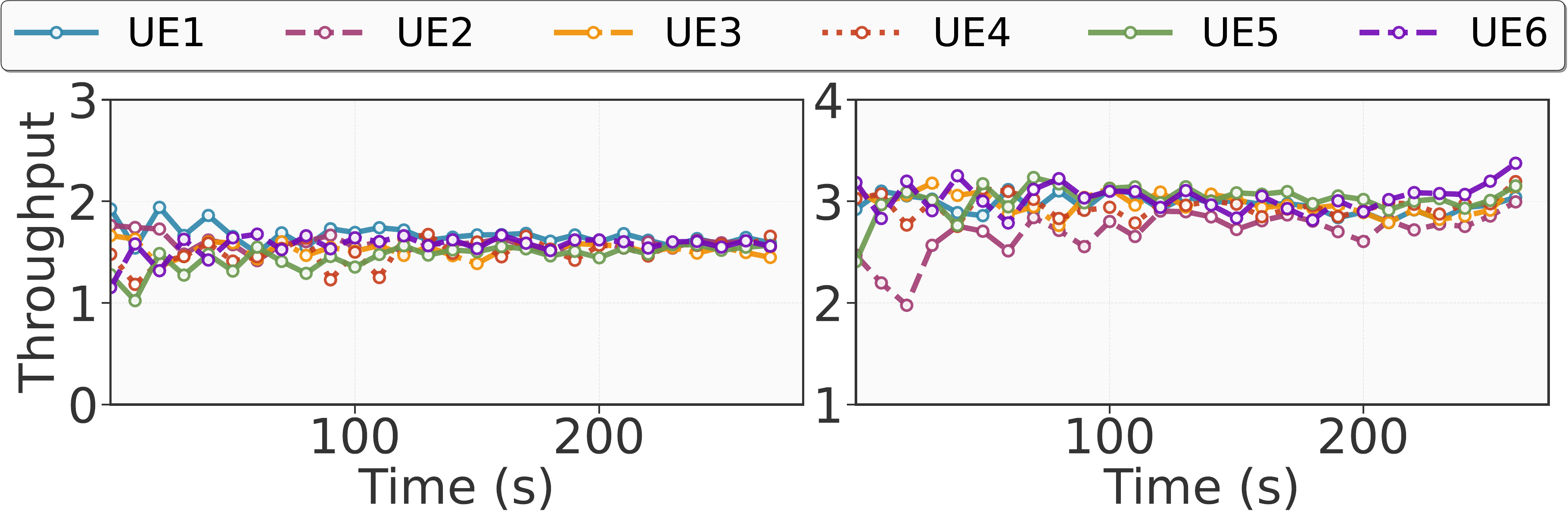}
    \tightcaption{Throughput (Mbps) for each file transfer application while running static (left) and dynamic (right) LC workloads.}
    \vspace{1em}
    \label{fig:file-transfer-performance}
\end{figure}

To verify that \sys{} does not starve BE traffic, we measure the average throughput of all BE UEs under both static and dynamic workloads (\autoref{fig:file-transfer-performance}). 
In both cases, BE UEs fairly share the remaining bandwidth, around 1.8 Mbps under static workload and 3 Mbps under dynamic workload, and no UE experiences prolonged starvation throughout the experiments.

\subsection{Impact of Edge Resource Schedulers}

\begin{figure*}[]
    \centering
    \begin{subfigure}[b]{0.48\textwidth}
        \centering
        \includegraphics[width=\textwidth]{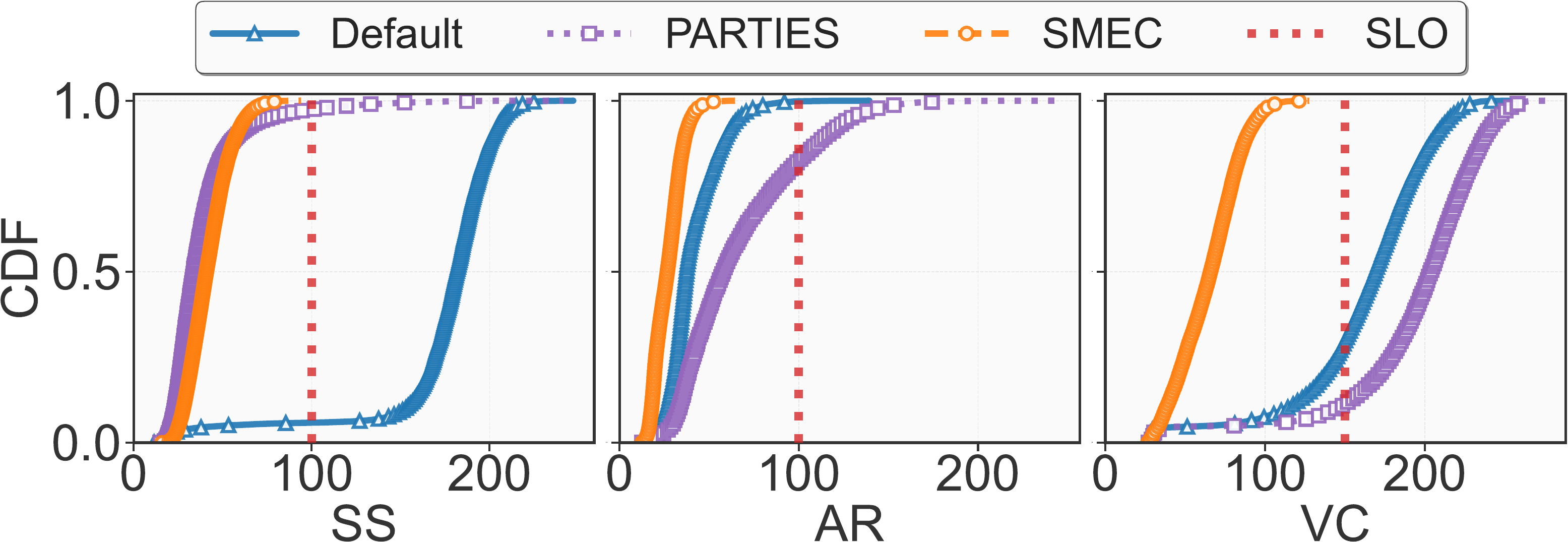}
        \caption{Static workload}
        \label{fig:proactive-vs-reactive-static-workload}
    \end{subfigure}
    \hfill
    \begin{subfigure}[b]{0.48\textwidth}
        \centering
        \includegraphics[width=\textwidth]{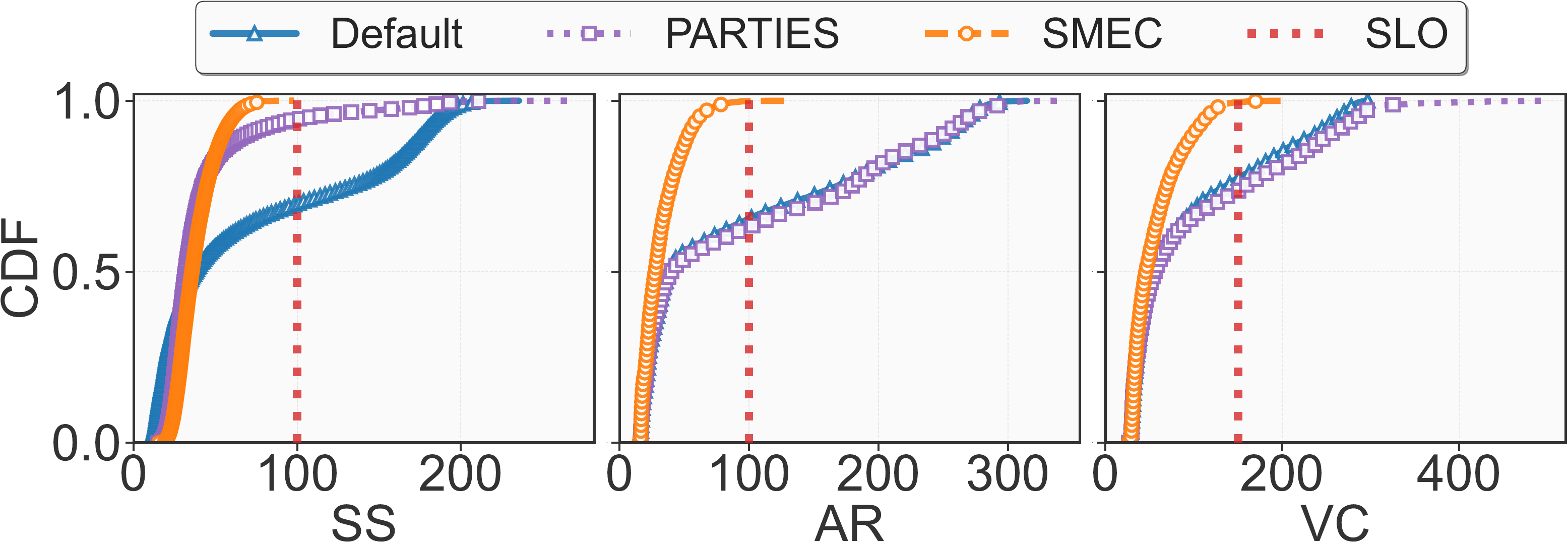}
        \caption{Dynamic workload}
        \label{fig:proactive-vs-reactive-dynamic-workload}
    \end{subfigure}
    \vspace{1em}
    \tightcaption{Processing latency (ms) at edge server with different resource schedulers.}
    \label{fig:proactive-vs-reactive-combined}
\end{figure*}

We evaluate the impact of different edge resource schedulers by comparing Default, PARTIES~\cite{parties:asplos19}, and \sys{}'s edge scheduler under static and dynamic workloads.
All experiments use \sys{}'s RAN scheduler to isolate edge effects, with processing latency as the primary metric (\autoref{fig:proactive-vs-reactive-combined}).
\label{sec:eval-server-schedulers}

\mypara{Static workload.}
\sys{} reduces SLO violations across all applications and lowers P99 processing latency by 1.5--4$\times$ \vs Default and PARTIES (\autoref{fig:proactive-vs-reactive-static-workload}).
PARTIES achieves reasonable performance for SS due to static workload but suffers from delayed feedback effects that inflate tail latency.
For AR and VC, PARTIES underperforms Default by sometimes prioritizing both LC applications simultaneously, amplifying GPU interference.

\mypara{Dynamic workload.}
\sys{} maintains superior SLO satisfaction and 2--4$\times$ lower P99 latency than baselines (\autoref{fig:proactive-vs-reactive-dynamic-workload}).
PARTIES shows 10\% more SLO violations compared to static workload due to increased variability.
Default and PARTIES perform poorly for AR and VC because they lack deadline awareness and accurate early drop, causing queue buildup and widespread SLO violations during bursts.

\subsection{Microbenchmarks}
\label{sec:eval-micro-benchmarks}

To complement our end-to-end evaluation, we conduct a set of microbenchmarks to evaluate the effectiveness of key components of \sys{}.

\subsubsection{Accuracy of Request Start Time Estimation} 
\label{sec:eval-micro-benchmarks-request-start-time-estimation}

\begin{figure}[]
    \centering
    \includegraphics[width=\columnwidth]{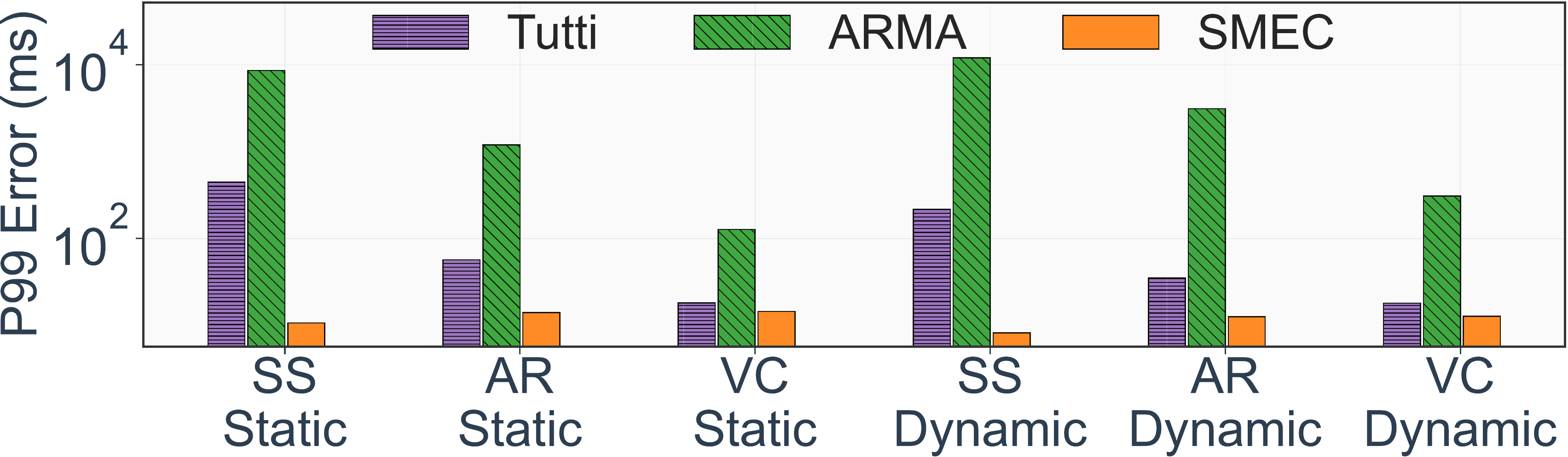}
    \tightcaption{Accuracy of request start time estimation.}
    \label{fig:remaining-time-accuracy}
\end{figure}

We evaluate the RAN scheduler's accuracy in estimating request start times using P99 absolute error (\autoref{fig:remaining-time-accuracy}). 
\sys{} achieves much lower error than \tutti{} and \arma{}. 
They infer request start times only after the edge server observes part of a request and notifies their RAN scheduler.
When uplink resources are not promptly allocated to LC applications, the server observes requests much later, and the notification to the RAN scheduler is delayed, causing large start-time estimation errors.
For example, for SS, \arma{} shows \second{10} of P99 error in both static and dynamic workloads, while \tutti{} incurs hundreds of milliseconds of P99 error.
By contrast, \sys{} relies on 5G control messages without any coordination with the edge server, making its estimation accurate and independent of uplink delays with only \ms{10} of P99 error.

\subsubsection{Accuracy of Latency Estimation}
\label{sec:eval-micro-benchmarks-network-and-processing-time-estimation}

\begin{figure}[]
    \centering
    \begin{subfigure}[b]{0.48\columnwidth}
        \centering
        \includegraphics[width=\textwidth]{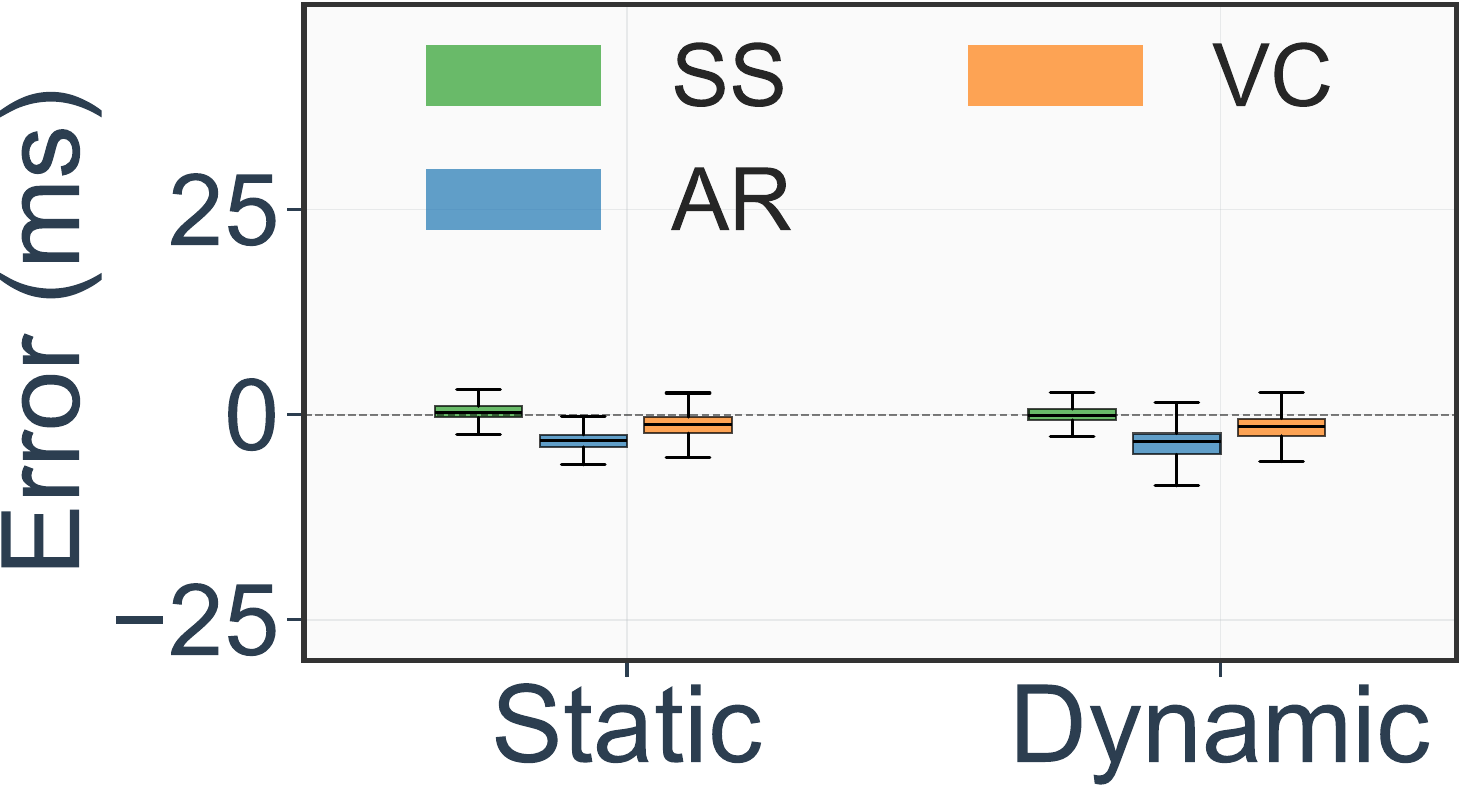}
        \caption{Network latency estimation}
        \label{fig:network-estimation-accuracy}
    \end{subfigure}
    \hfill
    \begin{subfigure}[b]{0.48\columnwidth}
        \centering
        \includegraphics[width=\textwidth]{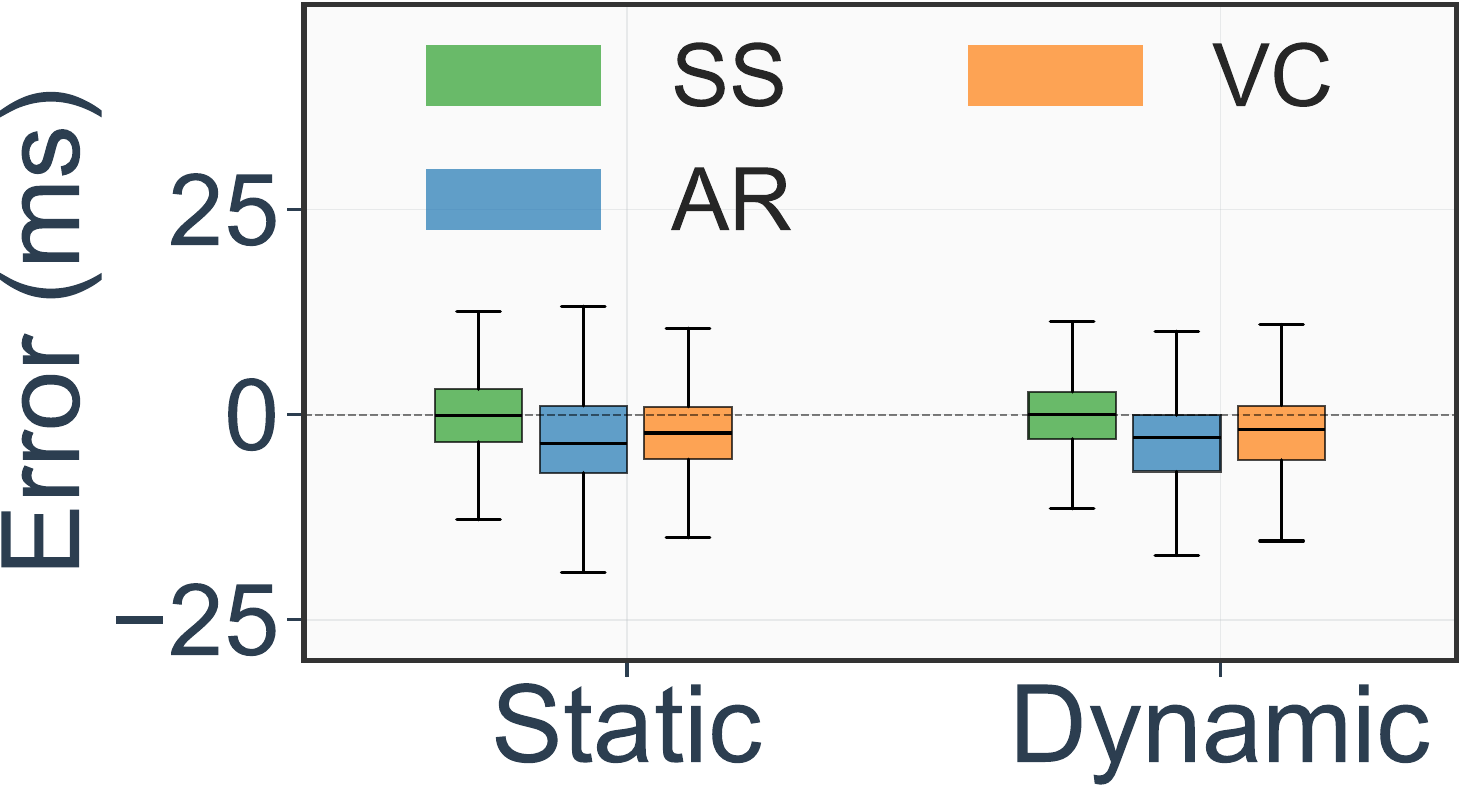}
        \caption{Processing time estimation}
        \label{fig:processing-time-estimation-accuracy}
    \end{subfigure}
    \vspace{1em}
    \tightcaption{Accuracy of network and processing latency estimation.}
    \label{fig:estimation-accuracy}
\end{figure}

We evaluate the accuracy of network and processing time estimation, which directly determines scheduling precision (\autoref{fig:estimation-accuracy}). 
\textit{\textbf{Network latency prediction}} is highly accurate, with errors typically within \ms{5} across all LC applications (\autoref{fig:network-estimation-accuracy}).
The error mainly stems from downlink transmission latency differences between ACK packets and actual responses. 
Our compensation mechanism (\autoref{sec:reference-time}) mitigates part of this difference, but a small residual error remains.
\textit{\textbf{Processing time estimation}} is also sufficiently accurate, with most requests showing errors within \ms{10} (\autoref{fig:processing-time-estimation-accuracy}). 
Larger errors stem from inherent per-request variance (\eg key frames in SS, complex scenes in AR) and amplified variance under compute contention.
Despite these errors, the prediction accuracy ensures that the majority of requests complete within their SLOs.

\subsubsection{Effect of Early Drop}
\label{sec:eval-micro-benchmarks-early-drop}

\begin{figure}[]
    \centering
    \includegraphics[width=\columnwidth]{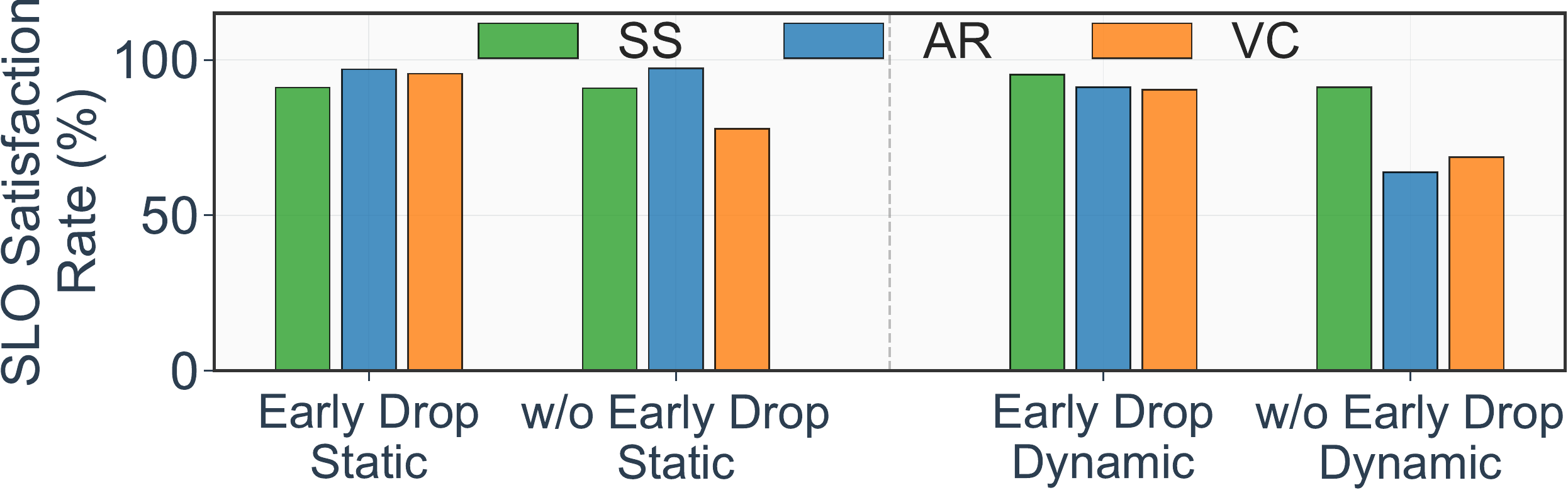}
    \tightcaption{SLO satisfaction rate with and without early drops.}
    \vspace{-1em}
    \label{fig:smec-drop-performance}
\end{figure}

Lastly, we evaluate the effectivenss of the early drop mechanism by measuring SLO satisfaction under both static and dynamic workloads (\autoref{fig:smec-drop-performance}). 
Early drop consistently improves performance across both workloads.
Under the static workload, where bursts are rare and mainly come from VC's sporadic traffic, early drop provides around 18\% points improvement primarily for VC.
Under the dynamic workload, where overloads are frequent, especially for GPU-heavy AR and VC, early drop brings over 20\% points improvement by discarding overly-urgent requests, freeing resources for requests likely to meet their deadlines. 

%% file: 08_discuss.tex
\section{Discussion}
\label{sec:discuss}

\mypara{Admission control for poor wireless channel conditions.}
While \sys{}'s RAN resource management is independent of wireless channel status, poor channel conditions can still degrade performance. 
LC applications with high bandwidth demands but weak channel quality may consume nearly all wireless resources while missing their SLOs, degrading performance for other users. 
An admission control mechanism can address this by profiling application throughput requirements against UE channel status and terminating service when channel quality is insufficient. 
This preserves SLO satisfaction for UEs with acceptable channel conditions while maintaining efficient spectrum utilization. 
Recent work such as Zipper~\cite{zipper24nsdi} provides techniques for designing such mechanisms.

\mypara{Handling downlink contention.}
Although this work focuses on uplink RAN scheduling, downlink contention is also important to consider. 
Downlink transmissions may suffer from congestion, but the downlink channel is usually more stable and predictable than uplink.
This stability enables end-to-end rate control mechanisms to converge to appropriate sending rates when timely client feedback is available. 
Prior work~\cite{zhuge22sigcomm} leverages uplink ACKs for faster congestion feedback, but delayed uplink grants in 5G networks can hinder timeliness. 
By improving uplink scheduling, \sys{} can prevent excessive queueing at UE buffers that would delay feedback packets (\eg TCP ACKs), ensuring timely congestion signals and enabling more effective downlink rate adaptation.

\mypara{Dealing with UE handover.}
While the applications studied in this work do not involve UE mobility across cells, scenarios such as autonomous driving may trigger inter-cell handovers. 
In deployments where multiple cells are managed under the same base station, handovers can often be absorbed locally without session disruption. 
The more challenging case involves mobility across different base stations, which requires transferring both radio and edge-session state across sites.
To support LC applications under such conditions, we envision proactively replicating UE state across base stations to enable seamless session continuation after handover. 
Designing such mechanisms introduces open questions around synchronization overhead and consistency.

\modified{\mypara{Fairness vs.\ latency trade-off.}
\sys{} prioritizes LC requests to meet their SLOs, consistent with our design goals (\autoref{sec:design-goals}).
Stronger throughput fairness between LC and best-effort (BE) traffic would delay LC transmissions under contention, increasing tail latency and reducing SLO satisfaction.
Our evaluation reflects this tension: systems that prioritize fairness between LC and BE traffic (\eg \tutti{} and \arma{}) incur more LC SLO violations than \sys{}, as shown in \autoref{fig:network-latency-steady-workload} and \autoref{fig:network-latency-dynamic-workload}.
Accordingly, for BE workloads \sys{} provides starvation freedom rather than strong throughput fairness under heavy contention.
Achieving strong fairness guarantees while preserving tight latency bounds for LC applications under wireless contention remains an open problem.

\mypara{Limitations.}
Our request identification approach at the RAN is constrained by the aggregation semantics of BSR.
When requests arrive in rapid succession within a single BSR reporting interval, they appear as a single aggregated increase at the MAC layer.
In such cases, \sys{} can only infer a shared request start time and performs scheduling at the request-group level rather than achieving per-request granularity.
This limitation affects scenarios with extremely high request rates but does not impact typical workloads where inter-request intervals exceed BSR reporting periods.

A second limitation concerns processing time estimation for compute resource scheduling.
\sys{} relies on processing time estimates to proactively allocate compute resources, which works well for applications with stable processing pipelines like those in our evaluation.
However, applications with dynamic processing patterns (\eg adaptive VR rendering) can exhibit high variance in processing latency, making accurate estimation challenging.
In such cases, estimation errors may reduce the effectiveness of \sys{}'s compute-side scheduling, though RAN-side scheduling remains unaffected and continues to minimize network delays.}

%% file: 09_concl.tex
\section{Conclusions}
\label{sec:concl}

We introduce \sys{}, a practical SLO-aware resource management framework for 5G MEC that achieves predictable performance without sacrificing deployment practicality.
By leveraging standard 5G control signals and natural application behaviors through lightweight APIs, \sys{} enables fully decoupled, deadline-aware scheduling at the RAN and edge with minimal application modifications.
Our 5G MEC testbed evaluation demonstrates that \sys{} achieves 90--96\% SLO satisfaction versus under 6\% for existing approaches, reduces tail latency by up to 122$\times$, and ensures starvation freedom for best-effort applications.
Thus, \sys{} provides a practical foundation for 5G MEC deployments that reliably meet the stringent demands of latency-critical applications.

%% file: acks.tex
\section*{Acknowledgements}
We would like to thank the anonymous NSDI reviewers and our shepherd, Timothy Wood, for their insightful comments and constructive feedback.
We also thank Xin Yuan, Li Shen, and Peirui Cao for their assistance with the measurements in this paper.
This work is based upon work supported by U.S. National Science Foundation (NSF) Awards 2403026 and 2326576.
Xiao Zhang was also supported by the Amazon AI PhD Fellowship.

%% file: 10_appendix.tex
\clearpage
\appendix

\section{Additional Measurement Results from Commercial MEC Deployments}
\label{appnd:measurements}

This appendix presents additional measurement results from our commercial MEC deployment studies that complement the findings presented in \autoref{sec:motivation}.

\subsection{End-to-End Latency of Augmented Reality }
\label{appnd:ar}

\begin{figure}[h!]
    \centering
    \includegraphics[width=\columnwidth]{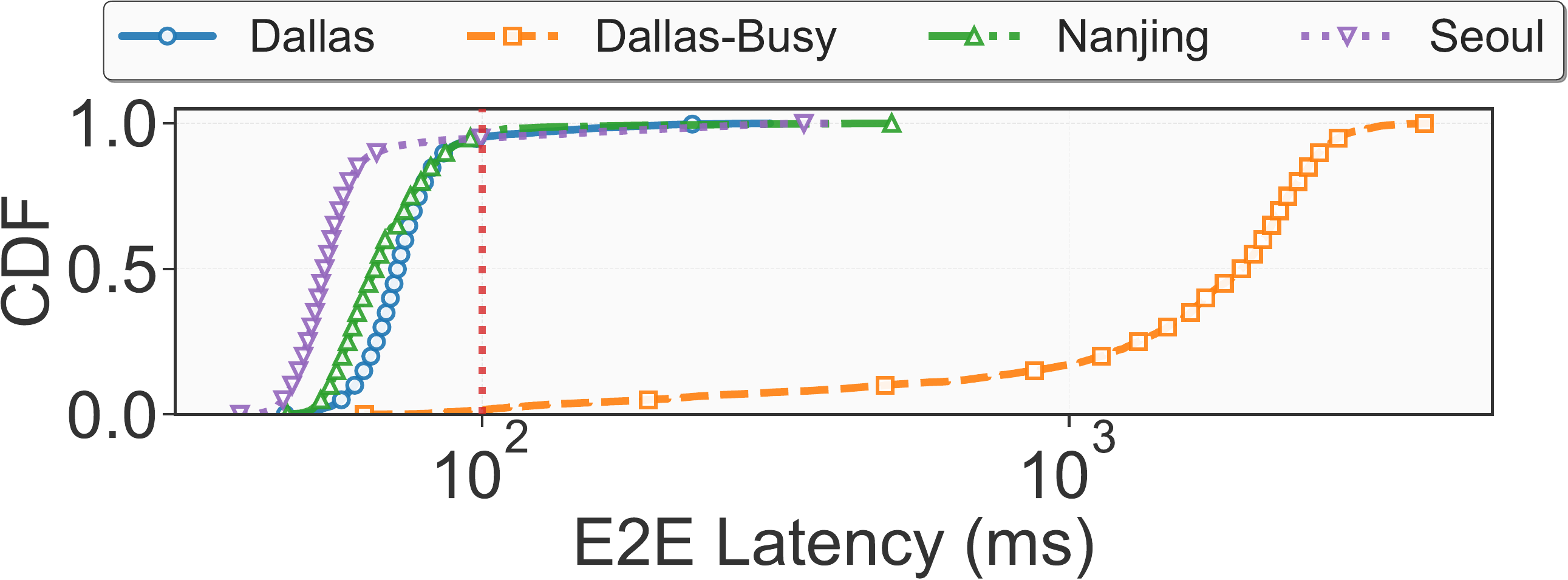}
    \caption{End-to-end latency for the augmented reality application without edge resource contention across MEC deployments in three cities. The dotted red line indicates the SLO.}
    \label{fig:motive-e2e-ar-ran}
\end{figure}

\autoref{fig:motive-e2e-ar-ran} shows that augmented reality (AR) exhibits a similar long-tail latency distribution as smart stadium during periods of low network activity. 
Since AR requires lower uplink throughput than smart stadium, only about \percent{5} of requests miss their SLO across all three cities. 
However, under high network activity (\dallas-Busy), wireless resource contention causes over \percent{98} of requests to violate their SLO.

\subsection{Effect of Compute Contention on End-to-End Latency}
\label{appnd:compute-contention}

We show the remaining results under compute resource contention for smart stadium in \nanjing and \seoul and augmented reality in all three cities.

\label{appnd:compute-contention-ss}
\begin{figure}[h!]
    \centering
    \includegraphics[width=\columnwidth]{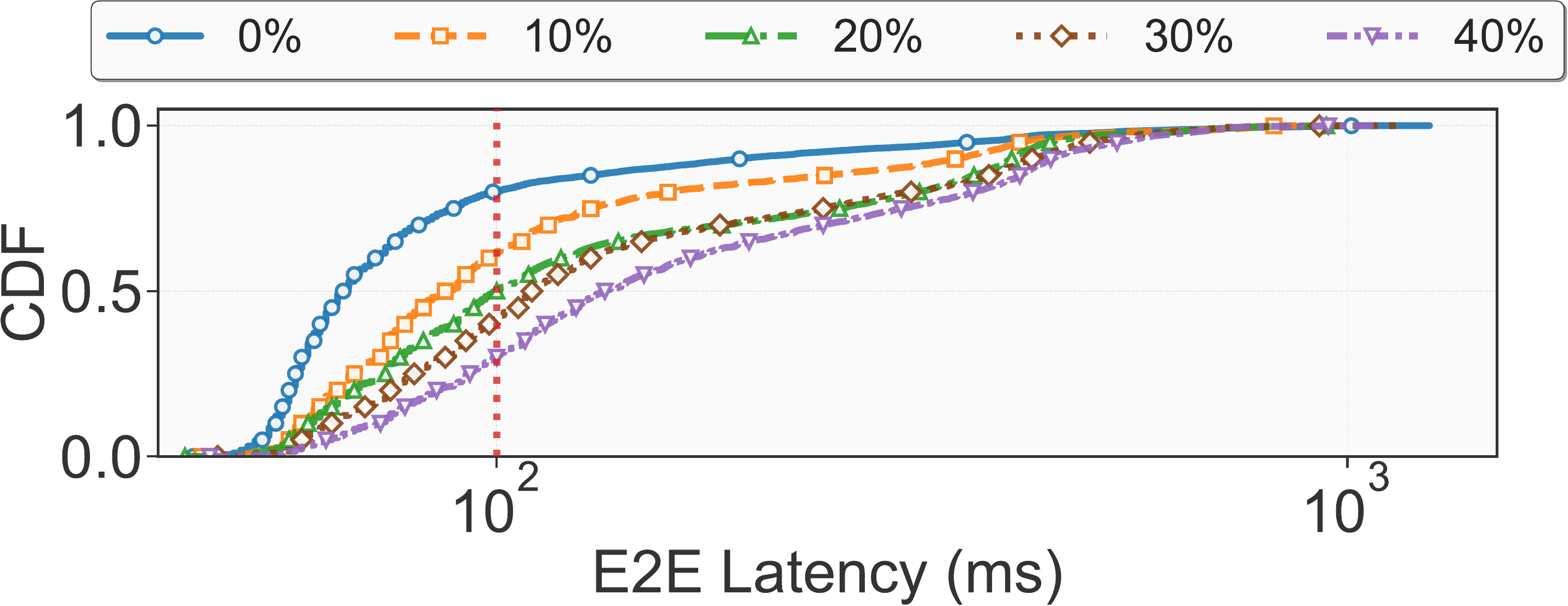}
    \vspace{-2em}
    \caption{End-to-end latency for smart stadium under different levels of compute resource contention in \nanjing. The dotted red line indicates the SLO.}
    \label{fig:motive-e2e-transcoding-compute-nanjing}
\end{figure}

\begin{figure}[h!]
    \centering
    \includegraphics[width=\columnwidth]{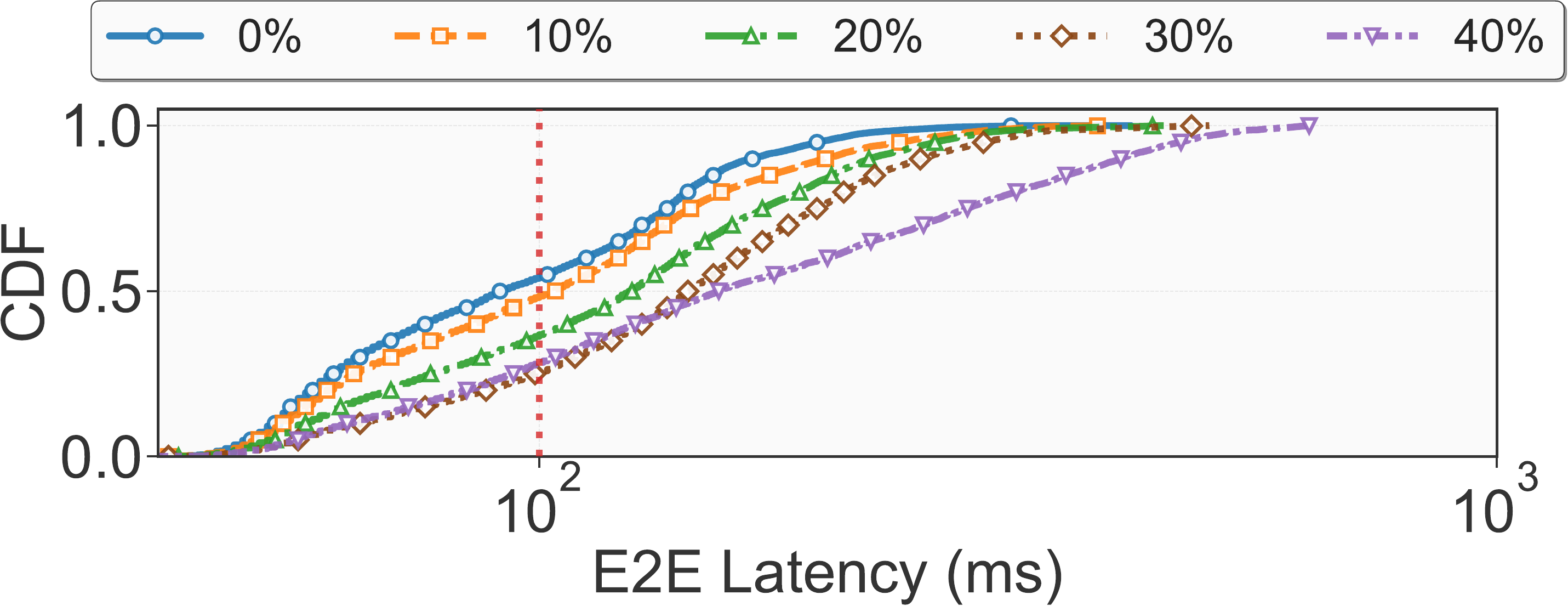}
        \vspace{-2em}
 
    \caption{End-to-end latency for smart stadium under different levels of compute resource contention in \seoul. The dotted red line indicates the SLO.}
    \label{fig:motive-e2e-transcoding-compute-seoul}
\end{figure}

\mypara{End-to-end latency of smart stadium.}
\autoref{fig:motive-e2e-transcoding-compute-nanjing} and \autoref{fig:motive-e2e-transcoding-compute-seoul} further confirm the effect of CPU resource contention on the end-to-end latency of CPU-intensive applications like smart stadium.
As CPU load increases, more requests exceed their SLO requirements and suffer from long tail latency.

\begin{figure}[h!]
    \centering
    \includegraphics[width=\columnwidth]{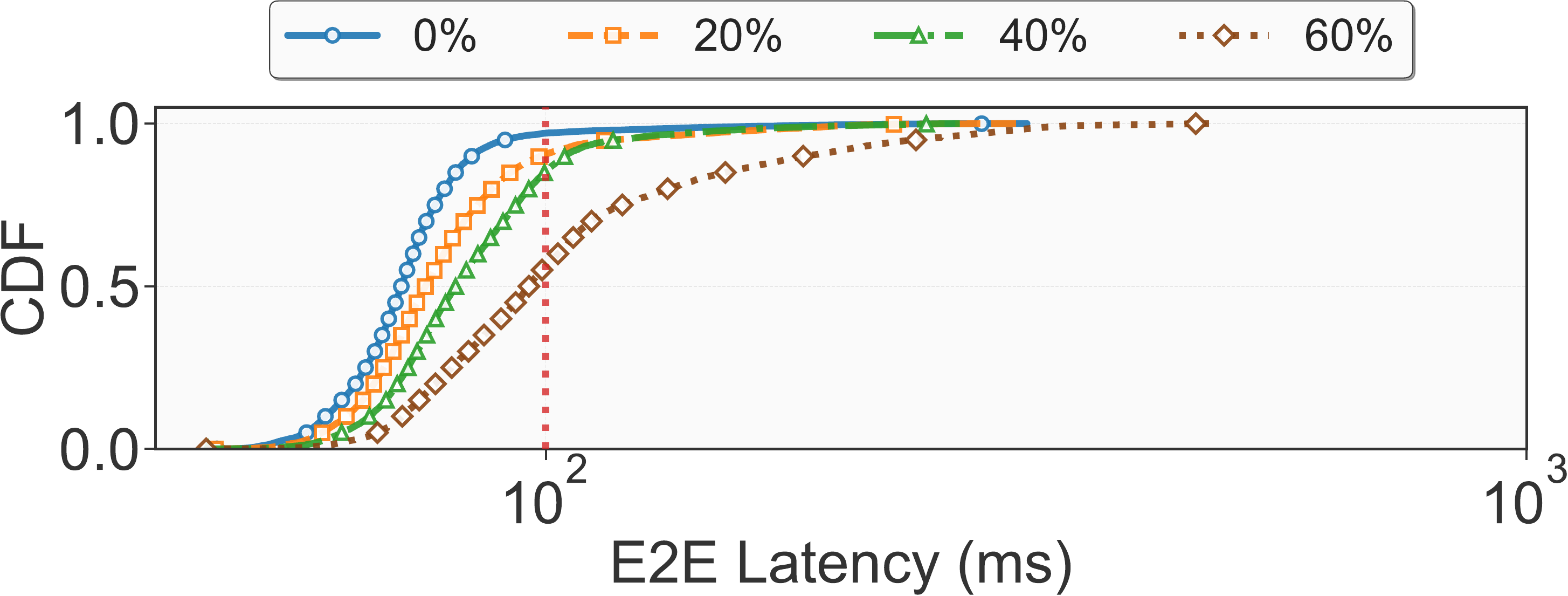}
        \vspace{-2em}

    \caption{End-to-end latency for augmented reality under different levels of compute resource contention in \dallas.}
    \label{fig:motive-e2e-ar-compute-dallas}
\end{figure}

\begin{figure}[h!]
    \centering
    \includegraphics[width=\columnwidth]{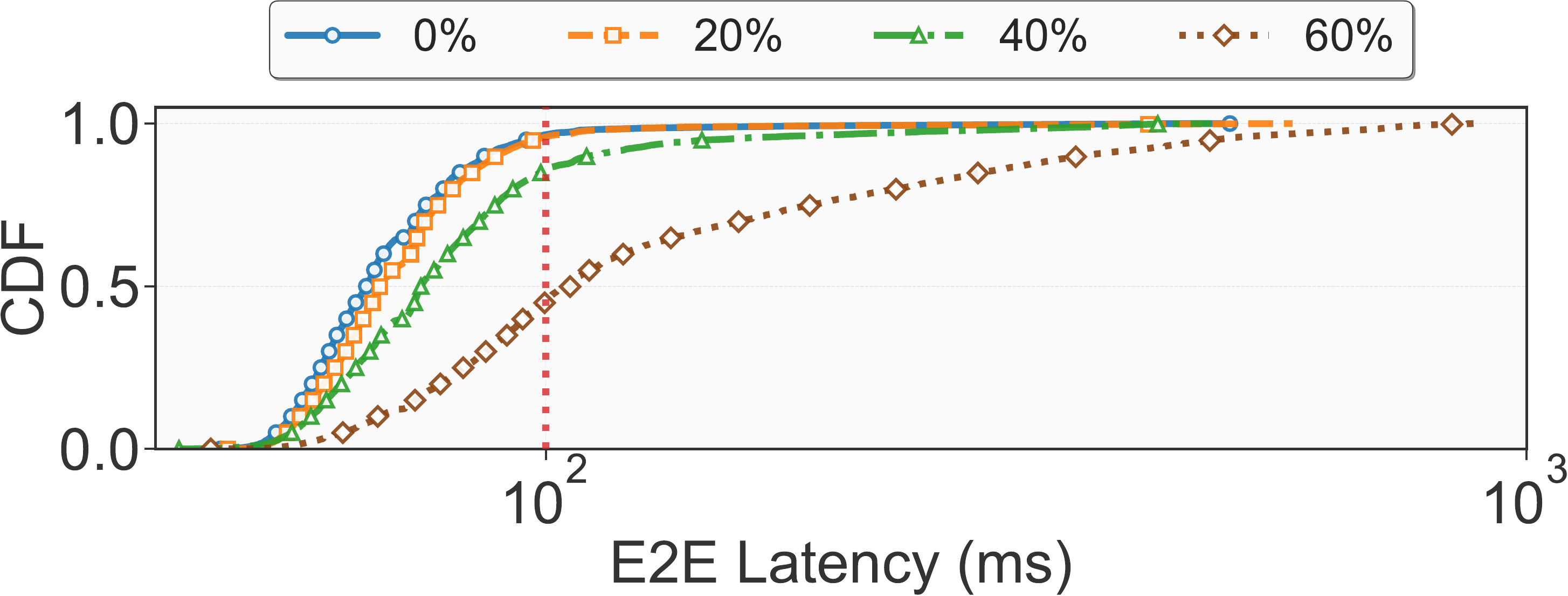}
    \vspace{-2em}

    \caption{End-to-end latency for augmented reality under different levels of compute resource contention in \nanjing.}
    \label{fig:motive-e2e-ar-compute-nanjing}
\end{figure}

\begin{figure}[h!]
    \centering
    \includegraphics[width=\columnwidth]{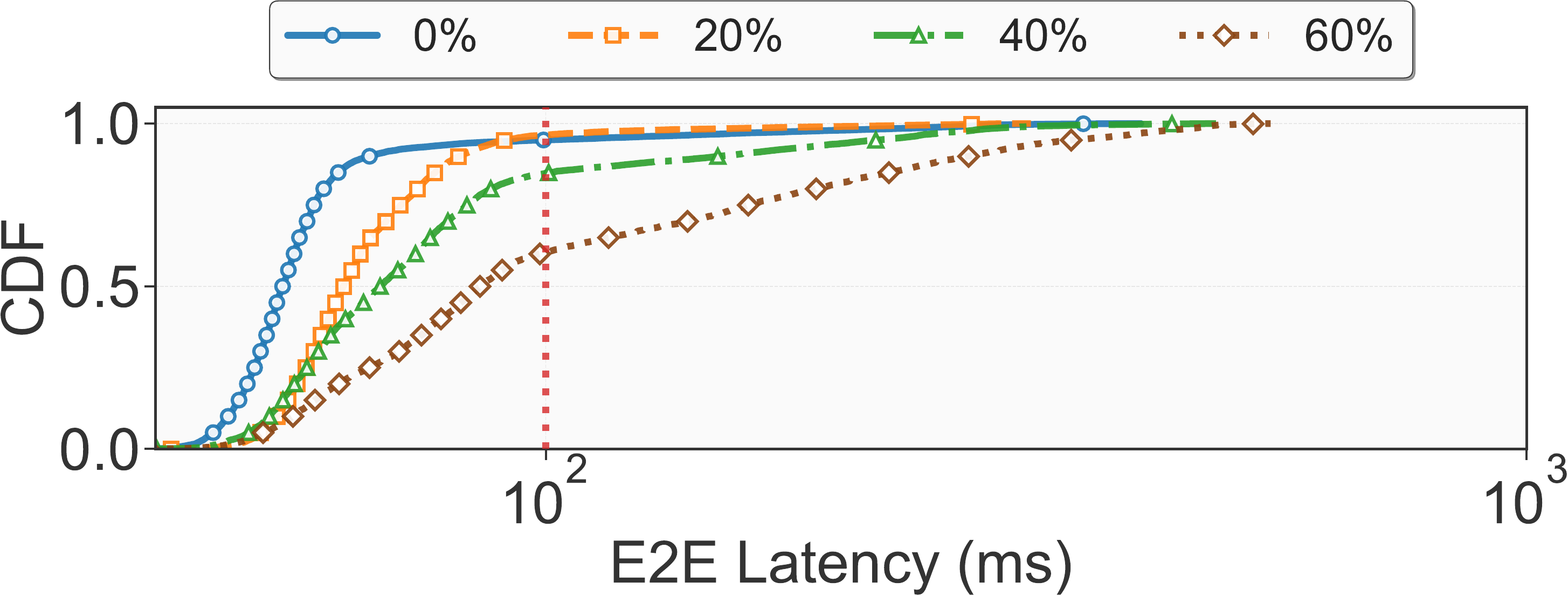}
    \vspace{-2em}

    \caption{End-to-end latency for augmented reality under different levels of compute resource contention in \seoul.}
    \label{fig:motive-e2e-ar-compute-seoul}
\end{figure}

\mypara{End-to-end latency of augmented reality.}
To further validate the impact of GPU contention on the end-to-end latency of GPU-intensive applications such as augmented reality, we run experiments in \dallas, \nanjing, and \seoul under varying levels of GPU load. 
We implement a GPU stressor using CUDA to emulate different contention levels. 
As shown in \autoref{fig:motive-e2e-ar-compute-dallas}, \autoref{fig:motive-e2e-ar-compute-nanjing}, and \autoref{fig:motive-e2e-ar-compute-seoul}, higher stress levels lead to more requests exceeding their SLOs and experiencing longer tail latencies.

\subsection{Variability of Network Latency} 
\label{appnd:latency-variability}

\begin{figure*}[]
    \centering
    \begin{subfigure}[b]{0.48\textwidth}
        \centering
    \includegraphics[width=\columnwidth]{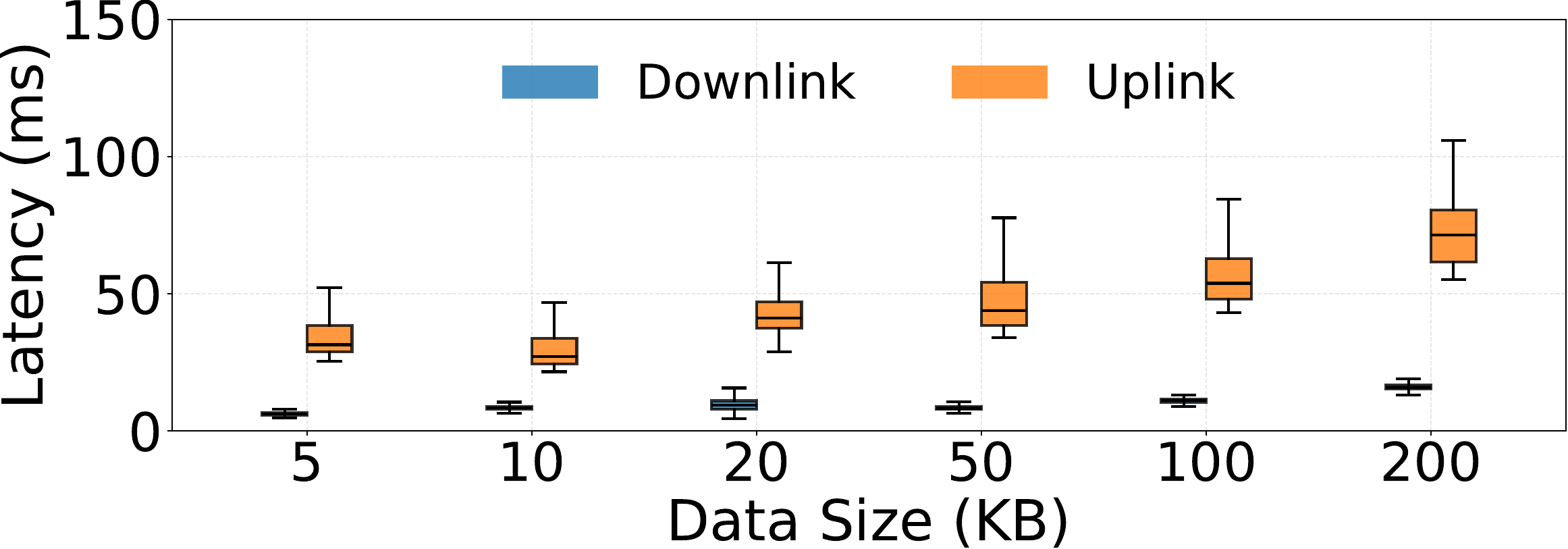}
    \caption{\nanjing}
    \label{fig:motive-decompose-nanjing}
    \end{subfigure}
    \hfill
    \begin{subfigure}[b]{0.48\textwidth}
        \centering
    \includegraphics[width=\columnwidth]{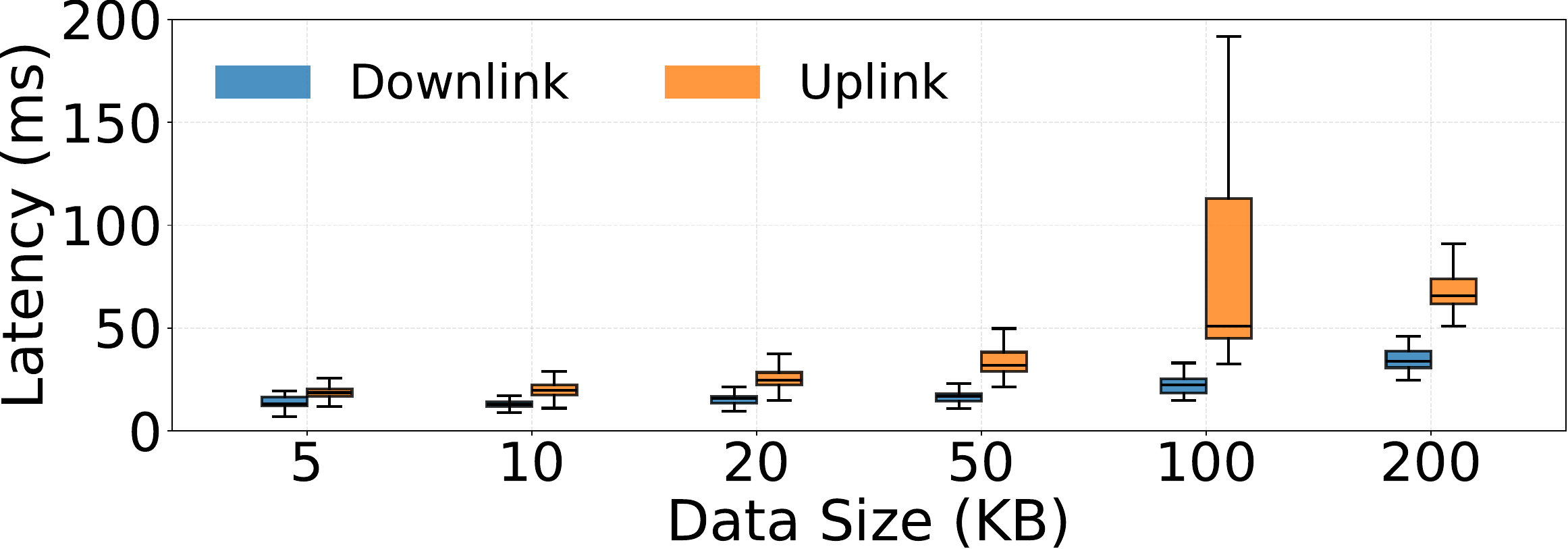}
    \caption{\seoul}
    \label{fig:motive-decompose-seoul}
    \end{subfigure}
    \caption{Network latency variability for uplink and downlink transmissions across different data sizes in \nanjing and \seoul.}
    \label{fig:motive-decompose-combined}
\end{figure*}

\begin{minipage}{\columnwidth}
\autoref{fig:motive-decompose-combined} reveals consistent asymmetry in both \nanjing and \seoul. 
As request size grows, uplink latency shows pronounced variability, whereas downlink latency remains largely stable. 
This resembles our observation that uplink paths suffer from high variability, while downlink paths maintain relative stability.
\end{minipage}

\section{Deadline-aware Proactive Edge Resource Scheduling Algorithm}
\label{appnd:edge-scheduling-algo}

\begin{algorithm}[htbp]
\small
\DontPrintSemicolon
\caption{\modified{Deadline-aware proactive edge resource scheduling.}}
\label{alg:edge-proactive-sched}
\modified{\textbf{Notation:} $r$ denotes a request; $a$ denotes the application that serves $r$; $\texttt{type}(r) \in \{\textsc{cpu}, \textsc{gpu}\}$.\;}

\modified{\BlankLine
\If{$t^{edge}_{budget}(r) \le 0$}{
\texttt{drop}(r)\;
\Return\;
}

\BlankLine
$urgency(r) \leftarrow t^{edge}_{budget}(r) / SLO_a$\;

\BlankLine
\eIf{$\texttt{type}(r)=\textsc{cpu}$}{
    \If{$urgency(r) < 0.1$}{
        \If{\texttt{now} $- t_{last\_cpu\_alloc}(a) \ge \ms{100}$}{
        \texttt{assign\_one\_more\_core}(a)\;
        $t_{last\_cpu\_alloc}(a) \leftarrow \texttt{now}$\;
        }
    }
    \If{$\texttt{cpu\_util}(a) < 60\%$}{
    \texttt{reclaim\_one\_core}(a)\;
    }
}{
    $prio(r) \leftarrow \texttt{map\_urgency\_to\_prio}(urgency(r))$\;
    \texttt{set\_cuda\_stream\_priority}(r,\ prio(r))\;
}

\BlankLine
\texttt{process\_request}(r)\;
}
\end{algorithm}